\documentclass[twocolumn,showpacs,preprintnumbers,amsmath,amssymb]{revtex4}
\usepackage{graphicx}
\usepackage{dcolumn}
\usepackage{color}
\usepackage{bm}
\usepackage{amsmath,amssymb,graphicx}
\usepackage{epsfig}
\usepackage{enumerate}
\usepackage{array}
\usepackage{multirow}
\DeclareMathOperator{\csch}{csch}

\begin{document}
\title{Environment dependent vibrational heat transport in molecular Junctions : Rectification, quantum effects, vibrational mismatch}
\author{Jayasmita Behera and Malay Bandyopadhyay }
\affiliation{SBS, I.I.T. Bhubaneswar, Argul, Jatni, Khurda, Odisha, India 752050.}

\vskip-2.8cm
\date{\today}
\vskip-0.9cm

\begin{abstract}
Vibrational heat transport in molecular junctions is a central issue in different contemporary research areas like Chemistry, material science, mechanical engineering, thermoelectrics and power generation. Our model system consists of a chain of molecules which sandwiched between two solids that are maintained at different temperatures. We employ quantum self-consistent reservoir model, which is built on generalized quantum Langevin equation, to investigate quantum effects and far from equilibrium conditions on thermal conduction at nanoscale. The present self-consistent reservoir model can easily mimic the phonon-phonon scattering mechanisms. Different thermal environments are modelled as (i) Ohmic, (ii) sub-Ohmic, and (iii) super-Ohmic environment and their effects are demonstrated for the thermal rectification properties of the system with spring graded or mass graded feature. The behavior of heat current across molecular junctions as a function of chain length, temperature gradient and phonon scattering rate are studied. Further, our analysis reveals the effects of vibrational mismatch between the solids phonon spectra on heat transfer characteristics in molecular junctions for different thermal environments.
\end{abstract}


\maketitle

\section{Introduction}
\label{intro}
\begin{figure}[t]
  \centering
  \includegraphics[width=\columnwidth]{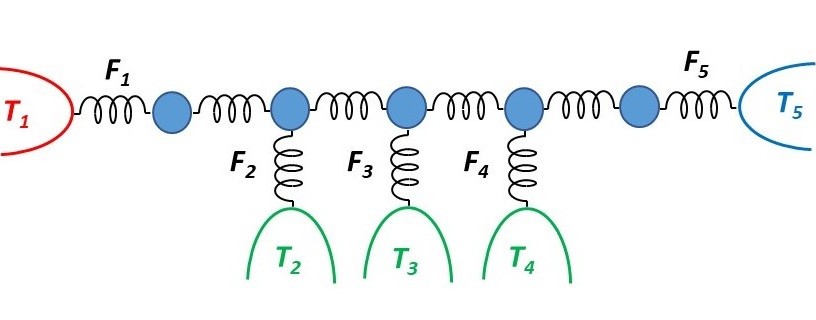}
  \caption{Scheme of a chain of $N=5$ particles; springs represent harmonic bond. The inner particles are connected to SC baths. The temperatures $T_{1}$ and $T_{5}$ sets the boundary conditions; the inner baths temperatures $T_{l}$ are determined by demanding the leakage current $F_{l}=0$. The net heat current flowing across the system is given by $F_{1}=-F_{5}$.}\label{fig1}
\end{figure}

\begin{figure}[t!]
  \centering
  \includegraphics[width=\columnwidth]{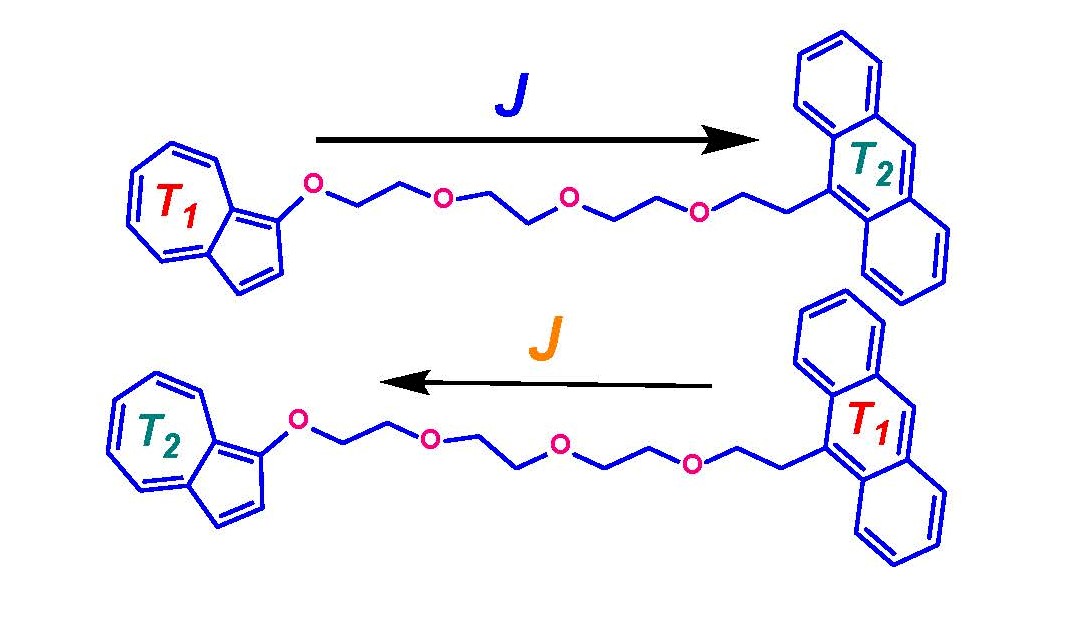}
  \caption{Heat current flow through $PEG_{4}$ oligomer bridging
two moieties (heat baths) for forward and reverse temperature gradient. Here $T_{1}>T_{2}.$}\label{alkane}
\end{figure}

In recent days a great amount of interest have been observed in the emerging field of molecular electronics with a focus on thermal transport through molecular junctions (MJs) using theoretical and experimental tools [1-6]. The low dimensional heat transport via phonons is a topic of a significant technological importance [7-10]. Typically our system consists of a chain of molecules bridging between two solids made of poorly electron-conducting components such as saturated hydrocarbons. Different atomic motions in the molecules (known as stretching mode) carry away thermal energy which they obtain through excited phonon modes in the contacts (solids) [11–13]. The focus of the present paper is thermal transport in molecules and different properties that control thermal conduction in molecular junctions [14-16]. This kind of studies have immense applications ranging from roughly  atoms to large biological molecules [17-18], thermoelectric devices [19], control of charge transfer [20-22] and photothermal processes in biological systems [23]. Understanding of vibrational heat transport can provide useful information about the chemical reactivity of protein folding dynamics [18]. Furthermore, understanding thermal energy transport in low dimensional systems is essential for developing electronic, mechanical, thermal, and thermoelectric devices, specifically organic–inorganic heterostructures [24-26]. \\
\indent
It is well known that quantum effects and anharmonic interactions play a vital role in vibrational heat transport. Particularly, these effects play a crucial role in thermal conductivity, thermalization process, development of Fourier's law, intra- and inter-molecular vibrational redistribution in short molecules, and in nonlinear and nonreciprocal effects [27-30]. But, it is a hefty task and a emerging research area to take into account both quantum effect as well as anharmonic interaction in the simulation of vibrational heat transport in molecular junctions. Classical molecular dynamics simulation can easily take care of the anaharmonic interaction but they are well above the quantum domain [31-33]. There are full quantum mechanical method based on Landauer formalism which only consider harmonic interaction [34-36]. Although nonequilibrium Green's function technique [37-40] and kinetic approaches can take into account anharmonic interactions but they are only applicable for systems with few quantum states or weak coupling limit or low anharmocity. On the other hand, the Born-Oppenheimer principle [41], mixed quantum-classical technique [42], numerically exact path integral technique [43] can handle only certain models like spin-Boson model.\\
\indent
Considering both the anharmonic interaction (arising from phonon-phonon scattering) and quantum effects, our analysis is based on quantum self-consistent reservoir model (QSCRM) to test the problem of vibrational heat transport in a chain of atoms bridging between two solids that are maintained at different temperatures. Our model is based on chains of $N$ beads and springs coupled to two solids at the edges, maintained at constant temperature differences. Each of the inner $(N - 2)$ beads are also connected to a self-consistent reservoir (SCR)
(see Fig.\,\ref{fig1}). In this way the inner (self-consistent) baths introduce an effective way to provide a simple scattering mechanism that might lead to local equilibration and to the onset of the (diffusional) Fourier's law of heat conduction [2,13]. On the other hand, the temperature of these $(N - 2)$ internal baths is determined by demanding that there is no net heat flow between the chain atoms and the reservoirs at steady state.\\
\indent
The classical version of the QSCRM was introduced
in Refs. [44-45]. Further, manifestation of Fourier’s law and development of local equilibrium in the thermodynamic limit are discussed in Refs. [46-47]. In the weak coupling limit the quantum version of
this model was studied by Visscher and Rich [48]. Beyond the weak coupling limit the quantum version of this model was considered in the linear response regime by Dhar and Roy [49,50]. Beyond the linear-response approximation, Bandyopadhyay and Segal introduced an iterative numerical procedure for the QSCRM to examine the phenomenon of thermal rectification in short MJs [51]. Heat transport in graphene nanoribbons were studied using massive atomistic simulations of QSCRM in Ref. [52]. Recently, Fereidani and Segal used QSCRM to study quantum effects and vibrational mismatch in the context of vibrational heat transport in short MJs [53].\\
\indent
It has been confirmed by simulations, time resolved vibrational spectroscopy, and theoretical investigations that energy transfer from one chemical group to another in one direction is more facile than in the reverse direction [54], even if both chemical groups are excited to similar energy. One of the issues of the present paper is to addressing thermal transport and rectification in a chain of molecules bridging between two heat baths which may have different structure of spectrum. As an illustrative example one may think about $PEG_4$ oligomer bridging two moieties, which serve as heat baths (see Fig. \ref{alkane}). The coupling between the oligomer and the Bosonic reservoir can be characterized by different forms of the reservoir spectrum for distinct physical scenarios. The three main classes that are usually considered in the literature are the so-called Ohmic, sub-Ohmic and super-Ohmic spectra [55]. We will compare the thermalization process and rectification process for these three types of reservoirs. Understanding which type of environment is most suitable for rectification and thermalization process can be of great importance in the choice of the physical system in building different molecular devices [56]. Heat transport via a quasi-one-dimensional system, like carbon nanotubes shows anomalous transport [57]. Thus, investigating the effect of different thermal environment using typical systems is an important tool not only for theoretical development but also for future development of molecular devices. \\
\indent
Different aspects of quantum effects and anharmonicity in the thermal
conduction of short molecules is demonstrated in a recent experiment [58]. In this experiment Majumdar et. al. considered a self-assembled monolayers of alkanes with about 10 methylene units which were sandwiched
between metal leads with distinct phonon spectra characterized by different Debye frequencies [58]. By considering different solids, they also found  that the thermal conductance decreased as the Debye-frequency mismatch between the two solids increased. Fascinated by this experimental results on phonon-mismatch and how quantum effects, anharmonicity, large temperature differences together with different thermal spectral densities play together to determine the thermal conductance of MJs --- is another focus of the present paper. \\
\indent
Although the radiative contribution (photonic contribution) of heat transport in nanoscale can not be ignored. But, this photonic contribution of heat transport is completely out of scope in the present paper. We mainly concentrate on the phononic heat transport at nanoscale.  But, one may go through Refs. [59-61] for recent developments in the radiative heat transport at nanoscale. \\
\indent
Having discussed this background, the rest of the paper is organized as follows. In Sec. II we present the
QSCR model and describe the Physical entity.
We further explain how to calculate the thermal properties of
the model in the classical limit, in the linear response limit and in the quantum regime.
Section III provides the heat current characteristics in different kind of systems, e.g., mass graded or spring graded domains, and manifest the onset of the thermal rectifying effect
in an asymmetric setting for three different thermal environments. In Sec. IV, we study the
thermal conductance of MJs with increasing molecular length, temperature, and phonon scattering rate for different spectral density of the environment. The combined effect of anharmonicity, quantumness, and phonon mismatch in thermal transport for the above mentioned three types of thermal environment are discussed in Sec. V. We discuss and conclude in Sec. VI.
\section{Model Hamiltonian and Formalism}
\label{section2}
In this section we describe our model system of linear chains which can represent oligomers, e.g.,
alkane chains [56], polyethylene glycol [62] molecules to calculate
the vibrational heat current across MJs utilizing the
QSCR technique. This technique is built on the generalized
quantum Langevin equation and the steady-state
properties can be calculated employing the Green’s function formalism [19,43]. To take into account far from equilibrium situations in our model for heat transport, we adopt the numerical
method introduced  in Refs. [51,53]. In the present analysis we go beyond the method discussed in  Refs. [51,53] and consider “structured” reservoirs with phonon
spectra that are categorized as (i) Ohmic, (ii) sub-Ohmic and (iii) super-Ohmic. For comprehensiveness of the readers,
we briefly describe the working equations of the QSCRM.
We consider a one-dimensional (1D) chain with $N$ particles connected by harmonic links. In the present paper we consider the problem of heat transfer in the quantum harmonic chain model with each inner site
connected to a self-consistent (SC) reservoir. This model
has been developed with the motivation to include nonlinear
behavior in an effective way [44-46].
Each particle is bilinearly connected to an independent thermal reservoir. The temperature of the reservoirs
attached to the first and last particles are set to $T_{1}$ and $T_{N}$, respectively. These two end reservoirs establish the boundary conditions and we refer them as ``physical" reservoirs. We refer the inner $(N-2)$ reservoirs as ``self-consistent" (SC) reservoirs which introduces scattering effects without energy dissipation. The temperature of the SC reservoirs $T_{l}$ ($l=2, 3,..., N-1$) are determined in a self-consistent manner by demanding that the net heat current between the chain atoms and the reservoirs is zero.\\
\indent
The Hamiltonian of the entire system consists of  chain (S), left reservoir (L), right reservoir (R) and the inner baths (I) is given by
harmonic potentials,
\begin{equation}\label{eq:1}
  H=H_{S}\,+H_{L}\,+H_{LS}\,+H_{R}\,+H_{RS}\,+\sum_{I}H_{I}\,+\sum_{I}H_{IS},
\end{equation}
where
\begin{equation}\label{eq:2}
 \begin{aligned}
  H_{S}&=\frac{1}{2}P_{S}^TM_{S}^{-1}P_{S}\,+\frac{1}{2}X_{S}^TK_{S}X_{S},\\
  H_{B}&=\frac{1}{2}P_{B}^TM_{B}^{-1}P_{B}\,+\frac{1}{2}X_{B}^TK_{B}X_{B},\\
  H_{BS}&=X_{S}^TK_{SB}X_{B}, \indent B=L, R, I.
 \end{aligned}
\end{equation}
We use the index $B = L, R, I$ for different reservoirs and internal baths. The real diagonal matrices $M_{S}$ and $M_{B}$ represent the masses of the particles in the chain and in the reservoirs or baths, respectively. The force-constant matrices for the chain and reservoirs (baths) are given by the real symmetric matrices $K_{S}$ and $K_{B}$, respectively. The term $K_{SB}$ includes the force-constant
coefficients between the chain and the baths. The column vectors $X_{S}$ and $X_{B}$ are the Heisenberg operators of the particles' displacements about some equilibrium positions with corresponding momenta
operators $P_{S}$ and $P_{B}$, respectively. The displacement operators and momenta operators satisfy the usual commutation relations $[X_{l},P_{m}]=i\hbar\delta_{l,m}$ and $[X_{l},X_{m}]=[P_{l},P_{m}]=0$.
The Heisenberg equations of motion for the chain and the baths are given as follows:

\begin{equation}\label{eq:3}
  M_{S}\ddot{X}_{S}=-K_{S}X_{S}\,-K_{SL}X_{L}\,-K_{SR}X_{R}\,-\sum_{I}K_{SI}X_{I}.
\end{equation}
\begin{equation}\label{eq:4}
   M_{B}\ddot{X}_{B}=-K_{B}X_{B}\,-K_{SB}X_{S},\quad B=L, R, I.
\end{equation}\\
Further, if we only consider a 1D chain with nearest neighbor coupling, the Hamiltonian of the system is simplified as follows
\begin{equation}\label{eq:5}
H_{S}=\sum_{s=1}^{N}\frac{1}{2}m_{s}\dot{x}_{s}^2\,+\sum_{s=1}^{N-1}\frac{1}{2}m_{s}\omega_{0}^2(x_{s}-x_{s+1})^2,
\end{equation}
where, force constant between the particles is given by $m_{s}\omega_{0}^2$ and the displacement from equilibrium of the $s$th particle is $x_s$. \\
\indent
Since the entire system is harmonic, the dynamics of the system can be written in terms of the generalized Langevin equation. After solving the Heisenberg equations of motion of the reservoir (Eq.\,(\ref{eq:4})) and then plugging it back into Eq.\,(\ref{eq:3}) we obtain the generalized Langevin equation
\begin{equation}\label{eq:6}
\begin{split}
 m_{s}\ddot{x}_{s}(t) &= -m_{s}\omega_{0}^2\,[2x_{s}(t)-x_{s-1}(t)-x_{s+1}(t)] \\
 & -\int_{-\infty}^{t}d\tau\dot{x}_{s}(\tau)\Gamma_{s}(t-\tau)+\eta_{s}(t), s=1, 2,...,N.
 \end{split}
\end{equation}
The integral term in the equation (\ref{eq:6}) represents the dissipation of energy from the system to the environment and the last term is the fluctuating force of the bath acting on the system. The index $s$, labelling the friction kernel and the fluctuating force, can also be used to pin down the physical reservoirs or internal baths. The dynamical friction kernel can be written as $\Gamma_{s}(t-\tau)=2\gamma_{s}f(t-\tau)$, where $\gamma_s$ is the friction coefficient and $f(t-\tau)$ is memory function. We may consider $\Gamma_{s}(t-\tau)=2\gamma_{s}\delta(t-\tau)$ for structureless and memoryless baths. To analyze the effect of vibrational mismatch on heat transport we use structured bath spectrum as follows [63]:
\begin{equation}\label{eq:7}
  \Gamma_{s}(\omega)=\gamma_{s}\omega^{r-1}\exp\Big({-\frac{|\omega|}{\omega_{d,s}}}\Big),
\end{equation}
where, $r=1$, $r<1$ and $r>1$ are characterized as Ohmic bath, sub-Ohmic bath, and super-Ohmic bath, respectively.
For  baths without exponential cut-off one can use $\omega_{d}\rightarrow\infty$ and for the exponential cut-off we may choose $\omega_{d}\sim\omega_{0}$ with $m=1$. The frequency coefficient $\gamma_s$ has different role for the physical and the SC reservoirs. $\gamma_{L,R}$ represent the bond energy of the molecule to the physical solids. On the other hand, $\gamma_{I}$ ($I=2,3, ...,(N-1)$) denotes the effective anharmonicity in the system and it is usually interpreted as intramolecular vibrational relaxation rate. Usually the effect of phonon-phonon scattering is strong when $\gamma_{I}$ is large and it reduces to harmonic limit when $\gamma_{I}=0$. For simplicity, we consider constant value of $\gamma_{I}$. In the frequency domain, the dissipation and fluctuation force satisfy ($k_{B}=1$)
\begin{equation}\label{eq:8}
\begin{split}
& \frac{1}{2}\langle\eta_{s}(\omega)\eta_{m}(\omega^\prime)+\eta_{s}(\omega^\prime)\eta_{m}(\omega)\rangle \\
&=\frac{\hbar\omega}{2\pi}\Gamma_{s}(\omega)\coth\Big(\frac{\hbar\omega}{2T_{s}}\Big)\delta(\omega+\omega^\prime)\delta_{s,m}.
\end{split}
\end{equation}
The steady-state heat current at each bath can be obtained from the position-momentum correlation function [64]. The phonon heat current towards the $s$th bead from the attached bath is given by
\begin{equation}\label{eq:9}
\begin{split}
F_{s}= & \sum_{m=1}^{N}\int_{-\infty}^{\infty}d\omega\Gamma_{s}(\omega)\Gamma_{m}(\omega)\big|[G(\omega)]_{s,m}\big|^2\frac{\hbar\omega^3}{\pi}\\
& \times[f(\omega, T_{s})-f(\omega, T_{m})].
\end{split}
\end{equation}
This multi-terminal transport expression represents incoming
energy from the $s$th particle to any of the other terminals and is termed  as the “QSCR result.”
Here the phonon Greens' function, $G(\omega)$, is the inverse of a tridiagonal matrix with off-diagonal elements $-m_s\omega_{0}^2$, diagonal elements $2m_s\omega_{0}^2-m_s\omega^2-i\omega\Gamma(\omega)$, and $f(\omega,T)=\big[e^{\hbar\omega/T}-1\big]^{-1}$ denotes the Bose-Einstein distribution function at temperature T. The temperature profile across the system is obtained by demanding that there is zero energy leakage from the system toward each of the SC baths,
\begin{equation}\label{eq:10}
  F_{i}=0,\qquad i=2, 3, ..., N-1.
\end{equation}
Equation (\ref{eq:10}) gives $N-2$ nonlinear equations which can be solved numerically (see Appendix A for details). The roots of the equation give the temperatures of the SC baths. Substituting these temperatures into the expression for $F_{1}$(or equivalently, into $F_{N}$) gives the steady-state net heat current flowing across the system, $J=F_{1}=-F_{N}$. \\
\indent
The QSCRM is also applicable to obtain the behavior of heat current in the Quantum Linear-Response (QLR) regime and in the classical (C) domain.\\
\indent
In the linear response regime when the temperature difference along the chain is small, $(T_{L}-T_{R})/T_{a}\ll1$, where $T_{a}=(T_{L}+T_{R})/2$, the differences in Bose-Einstein function $(f_{s}-f_{m})$ in Eq.\,(\ref{eq:9}) can be Taylor expanded to $(T_{s}-T_{m})\times\partial f/\partial T_{a}$. The QLR expression for heat current is now reduces to a linear expression,
\begin{equation}\label{eq:11}
  \begin{split}
     F_{s}^{QLR}= & \sum_{m=1}^{N}\int_{-\infty}^{\infty}d\omega\Gamma_{s}(\omega)\Gamma_{m}(\omega)\big|[G(\omega)]_{s,m}\big|^2\frac{\hbar\omega^3}{\pi} \\
       & \times\frac{\hbar\omega}{4T_{a}^2}\csch^2\bigg(\frac{\hbar\omega}{2T_{a}}\bigg)(T_{s}-T_{m}).
  \end{split}
\end{equation}\\
Similarly, in the classical domain we consider high temperature limit, $f(\omega, T)\sim T/(\hbar\omega)$, thus Eq.\,(\ref{eq:9}) reduces to
\begin{equation}\label{eq:12}
  F_{s}^C=\sum_{m=1}^{N}\int_{-\infty}^{\infty}d\omega\Gamma_{s}(\omega)\Gamma_{m}(\omega)\big|[G(\omega)]_{s,m}\big|^2\frac{\omega^2}{\pi}(T_{s}-T_{m}).
\end{equation}\\
Since equations (\ref{eq:11}) and (\ref{eq:12}) are linear in SC baths temperatures, we can organize these equations as $F_{s}=\sum_{m} C_{s,m}(T_{s}-T_{m})$, where $C_{s,m}$ contains the frequency integration. The solution for the temperature profile of the SC bath, demanding $F_{s}=0$ for $s=2,3, ..., N-1$, is given by $\textbf{T}=A^{-1}v$ [49].
Here $A$ is a $(N-2)\times (N-2)$ diagonal matrix with diagonal elements $\sum_{m\neq s}C_{s,m}$ and nondiagonal elements $-C_{s,m}$. $v$ is a vector defined as $v_{s}=C_{s,1}T_{1}+C_{s,N}T_{N}$. The vector $\textbf{T}$ includes the steady-state temperatures of the inner baths. The net heat current $F_{1}$ can be readily calculate from the vector $\textbf{T}$.
\subsection{Physical Entity}
We now try to express our simulation results in terms of physical units. As we are interested to tally our results for alkane chains, we may consider a mode of frequency $1000$ cm$^{-1}$ for the single bond carbon-carbon stretching motion. This translates into a frequency of $\omega_0\sim 3\times 10^{13}$ Hz. The other parameters which are used in the simulation are $T_a\sim 1,\gamma_{L,R}=0.2, \gamma_I=0.8,$ and $\Delta T=0.8$. As a result one can obtain heat current $J\sim 0.02 - 0.04 $ for a chain length with $N \sim 6-12$ particles. Based on $\omega_0 \sim 3\times 10^{13}$ Hz, one can obtain $T_a=230$ K, $\Delta T = 180$ K (figure 4 of Ref. [56]), $\hbar\gamma_{L,R}= 4$ meV for the molecule-surface contact energy and $\hbar \gamma_I = 16$ meV for the vibrational energy of scattering or 25 ps for scattering rate. Thus the heat current translates as $J \sim 2\times 10^{-9} - 4\times 10^{-9}$ W. Furthermore, one may calculate the thermal conductance which is given by $ K_T=J/\Delta T= 14 - 20$ pW/K and it is comparable to experimental results on alkane chains with $4-20$ units [58]. \\
\indent
We furthermore consider Debye solids and other kind of environmental spectrum with
characteristic frequencies in the range of $0.1-1$, in units of $\omega_0$ and it
changes to $ \omega_d= 3\times 10^{12} – 3 \times 10^{13}$ Hz. If we consider
Debye temperature, these numbers actually translates to $T_D = 22 –
220$ K. Thus, this paper will enable us to implement a feasible technique
to simulate quantum heat transport in molecular junctions.
\section{Rectification}
\begin{figure}[b!]
  \centering
  \includegraphics[width=\columnwidth]{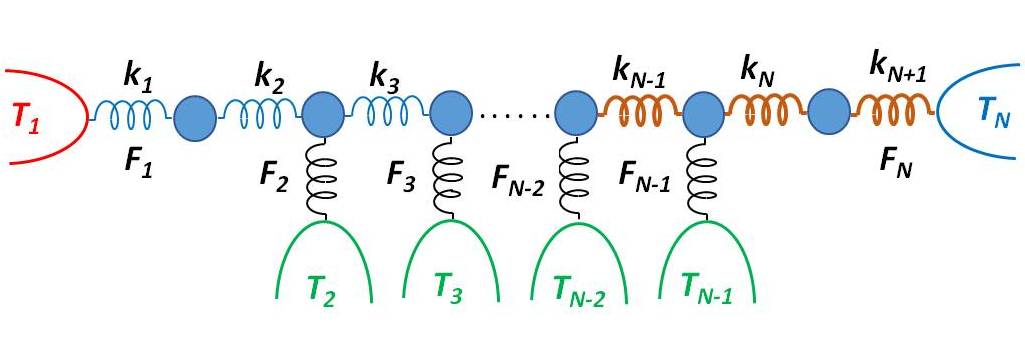}
  \caption{A scheme of Spring graded system with SC baths for $N$ particles. The spring graded system is described by dividing the system into two halves with both halves having different values of spring constants. Each half have identical value of spring constants. The two halves are shown by different color springs on both sides of the system separated by the dotted line.} \label{fig4}
\end{figure}

In the last decade, we have seen a large number of studies on thermal rectification which is nothing but an
 asymmetry of the heat current for forward and reverse temperature gradient [65]. It is usually seen that a well behaved rectifier acts as a heat conductor in one direction of temperature bias and it behaves as an insulator for the opposite direction of bias. It is well known that junctions which may incorporate an effective anharmonicity in the system with some sort of spatial asymmetry may demonstrate rectification [66]. Further, the self-consistent reservoirs include anharmonicity to the system in an effective way and it is useful to explore its rectification properties when we incorporate spatial asymmetry either by using spring graded chain (Fig.\,\ref{fig4}) or by using a mass graded chain (Fig.\,\ref{fig6}). The study on classical SC model and quantum linear-response (QLR) SC model prove that they will never show any kind of rectification even with the inclusion of spatial asymmetry [49]. Further studies reveal that full quantum treatment of the QSCR model exhibit rectification [51]. But the effect of structured bath on the heat rectification is still open.  In this section we study the rectification effect for both spring graded and mass graded chain systems for three different types of environment i.e., Ohmic, sub-Ohmic, and super-Ohmic environment. The bath spectrum is already defined in Eq.\,(\ref{eq:7}).
 We study the extent of rectification from the rectification ratio ($R$) which is defined  as follows
\begin{equation}\label{eq:16}
  R=\frac{J_{+}-J_{-}}{(J_{+}+J_{-})},
\end{equation}
where $J_{+}$ and $J_{-}$ are the absolute value of heat current for forward ($T_{1}>T_{N}$) and reversed ($T_{N}>T_{1}$) temperature biases, respectively.

\subsection{Spring graded system}
\begin{figure}[t!]
\centering
\includegraphics[width=\columnwidth]{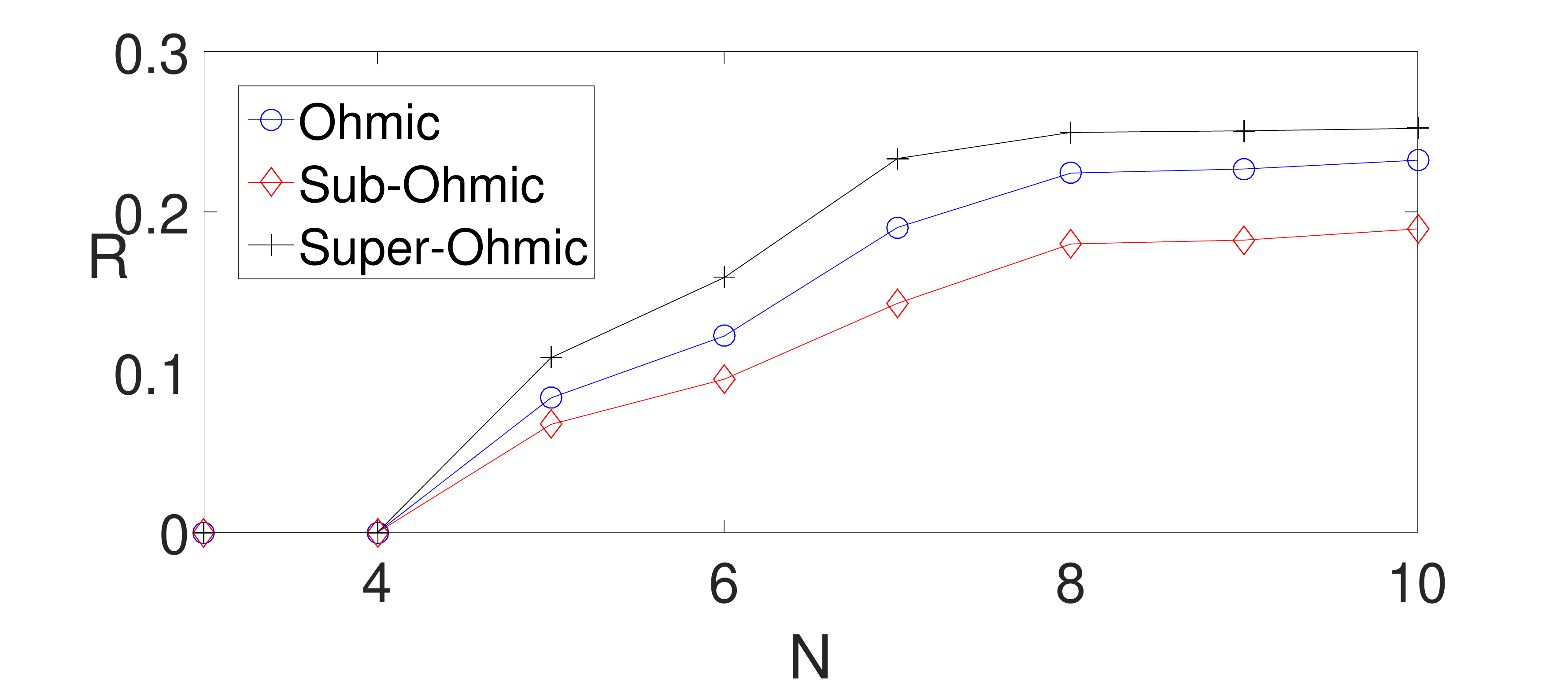}
\caption{Rectification in a spring graded system: Rectification ratio as a function of chain size for Ohmic (o), sub-Ohmic ($\diamondsuit$), and super-Ohmic (+) environments. Parameters are $\gamma_{0}=0.2$ for all sites, $k_{i}=1$, for $i=1,2, ..., 5$ and $k_{i}=4$ for $i=6,7,...,(N+1)$. In the forward direction we used $\beta_{1}=1,\beta_{N}=5$. Here $\beta\equiv1/T$ is the inverse temperature. The temperature difference
$\Delta T=T_{1}-T_{N}=0.8$. The average temperature $T_{a}=(T_{1}+T_{N})/2=0.6$.
The temperatures of left and right reservoir are interchanged in reverse direction.}\label{fig5}
\end{figure}
\begin{figure}[t]
  \centering
  \includegraphics[width=\columnwidth]{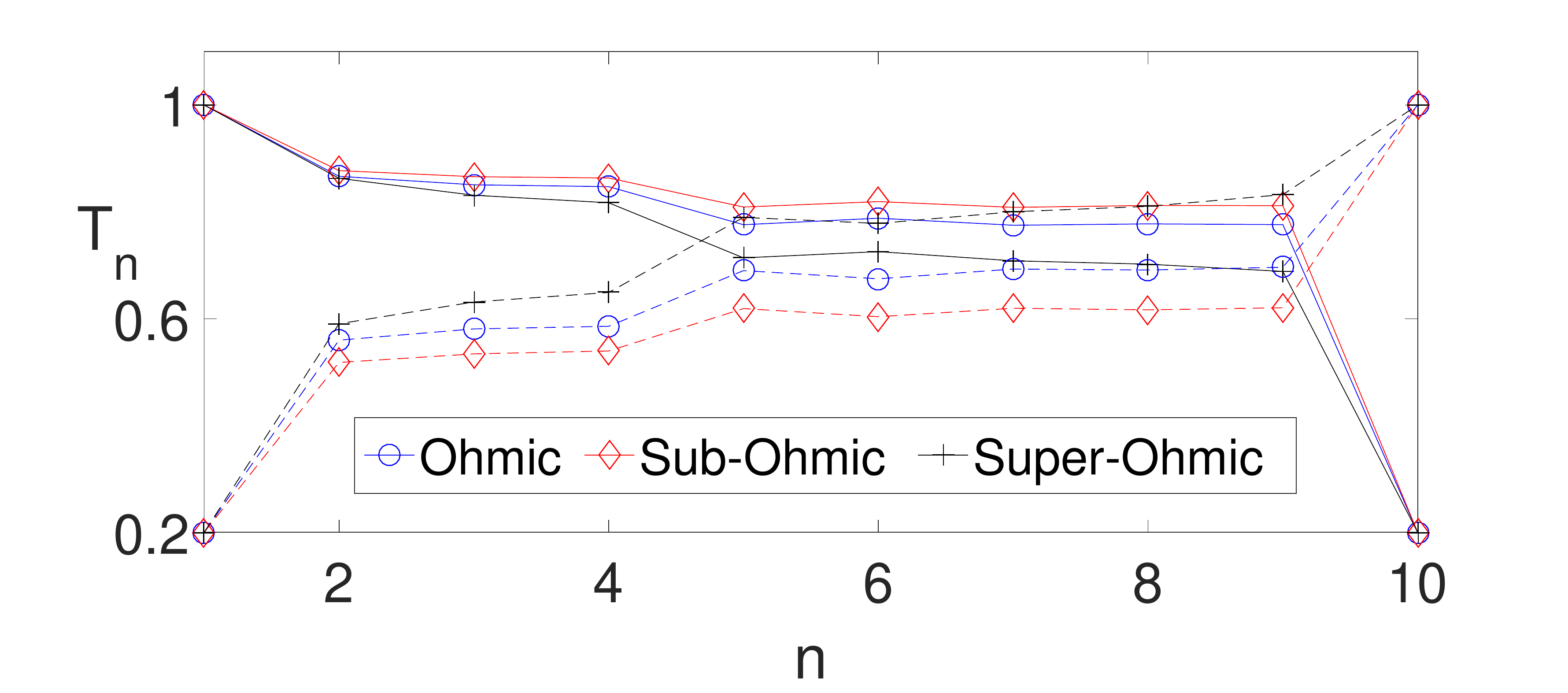}
  \caption{Temperature profile at site n for a spring graded $N=10$ chain in forward direction (solid line) and in reverse direction (dashed line) for Ohmic (o), sub-Ohmic ($\diamondsuit$), and super-Ohmic (+) environments. Parameters are $\gamma_{0}=0.2$ for all sites, $k_{i}=1$, for $i=1,2, ..., 5$ and $k_{i}=4$ for $i=6,7,...,(N+1)$. In the forward direction we used $\beta_{1}=1,\beta_{N}=5$. Here $\beta\equiv1/T$ is the inverse temperature. The temperature difference
$\Delta T=T_{1}-T_{N}=0.8$. The average temperature $T_{a}=(T_{1}+T_{N})/2=0.6$. The temperatures of left and right reservoir are interchanged in reverse direction. }\label{temp_k14}
\end{figure}
We use the QSCR method for the spring graded system with variable force constant to study the rectification effect. Using a non-equilibrium molecular dynamics (MD) simulation of single walled carbon nanotube (SWNT) people have investigated the defect-induced thermal rectification [67]. Hayashi et. al. [67] considered
vacancy defects only in half of the region along the axial direction. Their results confirm that heat flow from the defective SWNT to the pristine SWNT is nearly $10\%$ greater than in the opposite direction and it
 agrees with the experimental results of Wang et.al [68]. The defect is introduced by controlling the spring constant $k_i$. A defective SWNT has smaller number of $sp^2$ bonds which results in weaker interaction in the defect region with a smaller $k$ value than the pristine region. We describe the spring graded system by dividing the system into two halves with both halves having different values of spring constants. Each half have identical value of spring constants. For a system of N particles we take $k_{i}=1$ for $i=1,2, ...,N/2$ and $k_{i}=4$, for $i=N/2+1, ...,(N+1)$. For classical and QLR case the effect is negligible and one may observe $J_{+}\sim J_{-}$. On the other hand, for low temperatures and large bias, using QSCR method, one can demonstrate that  $J_{+}$ and $J_{-}$ deviate significantly. Figure \ref{fig5} displays the plot of rectification ratio $R$ with Chain size $N$ for Ohmic, sub-Ohmic and super-Ohmic environments. Here we study a chain of $N=10$ particles. We find that the rectification ratio increases with chain size and rectification is highest for super-Ohmic environment and least for sub-Ohmic environment.
\subsection{Mass graded system}
\begin{figure}[b]
  \centering
  \includegraphics[width=\columnwidth]{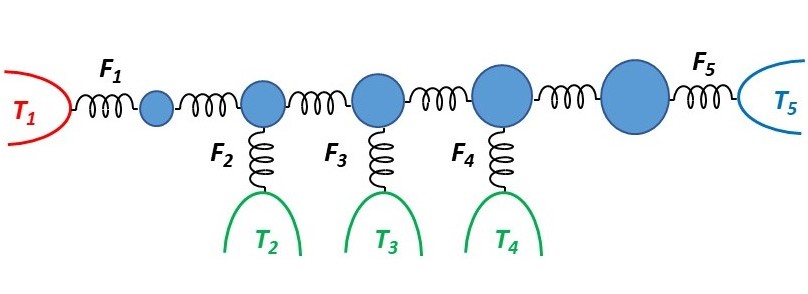}
  \caption{A scheme of mass graded system with SC baths for $N=5$ particles.}\label{fig6}
\end{figure}
\begin{figure}[b!]
  \centering
  \includegraphics[width=\columnwidth]{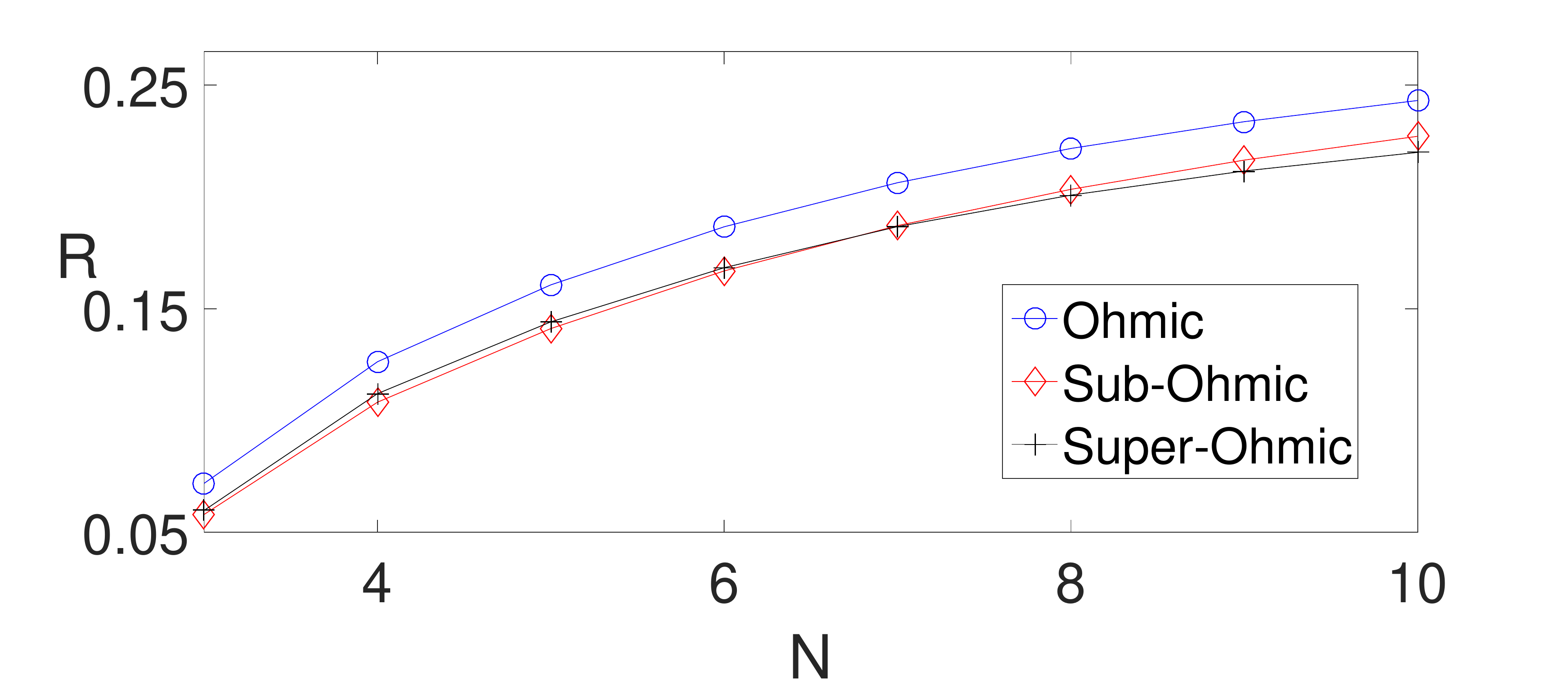}
  \caption{Rectification in a mass graded system: Rectification ratio as a function of chain size for Ohmic (o), sub-Ohmic ($\diamondsuit$), and super-Ohmic (+) environments, $\gamma_{0}=0.2$ for all sites, $M_{1}=0.2$, $M_{n}=M_{1}+(n-1)\times0.2$. In the forward direction we used $\beta_{1}=1,\beta_{N}=5$. Here $\beta\equiv1/T$ is the inverse temperature. The temperature difference $\Delta T=T_{1}-T_{N}=0.8$. The average temperature
   $T_{a}=(T_{1}+T_{N})/2=0.6$. The temperatures of left and right reservoir are interchanged in reverse direction.}\label{fig7}
\end{figure}
\begin{figure}[t]
  \centering
  \includegraphics[width=\columnwidth]{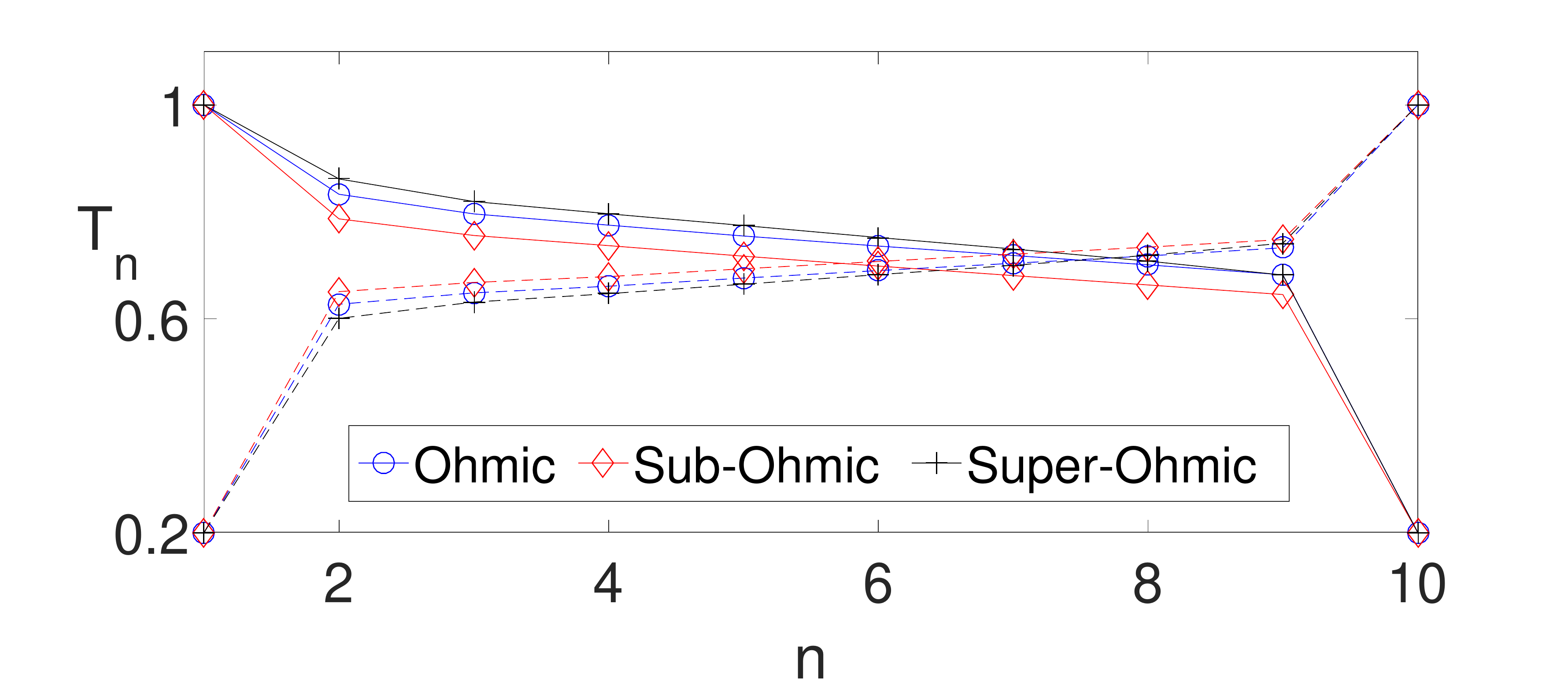}
  \caption{Temperature profile at site n for a mass graded $N=10$ chain in forward direction (solid line) and in reverse direction (dashed line)for Ohmic (o), sub-Ohmic ($\diamondsuit$), and super-Ohmic (+) environments, $\gamma_{0}=0.2$ for all sites, $M_{1}=0.2$, $M_{n}=M_{1}+(n-1)\times0.2$. In the forward direction we used $\beta_{1}=1,\beta_{N}=5$. Here $\beta\equiv1/T$ is the inverse temperature. The temperature difference $\Delta T=T_{1}-T_{N}=0.8$. The average temperature
   $T_{a}=(T_{1}+T_{N})/2=0.6$. The temperatures of left and right reservoir are interchanged in reverse direction.}\label{temp_mass}
\end{figure}
In the present subsection we demonstrate thermal rectification in an asymmetric quantum harmonic systems with SC baths by solving  the set of nonlinear equations (9).
We incorporate spatial asymmetry in QSCR model by using a mass graded system in which mass of the particles increases with the increase of chain size. However, as correctly pointed out in Ref. [69], most of the theoretical studies of heat rectification are  based on the sequential coupling
of two or three segments with different anharmonic potentials and they have following two drawbacks : (a) their experimental implementation are very much difficult and (b) the rectification power typically decays to zero with increasing the system size. Thus, people have found another way to produce the required asymmetry for thermal
rectification by considering a mass gradient along the
system length. This idea was first employed to experimentally
build a rectifying device with a carbon and boron-nitride nanotube
inhomogeneously mass-loaded with heavy molecules [70].  The experimental work with mass-loaded nanotubes was further numerically simulated (nonequilibrium molecular dynamics ) by means of a one-dimensional (1D) Fermi–
Pasta–Ulam (FPU) anharmonic oscillator lattice with a mass gradient [71-73].  Further theoretical studies on mass graded system can be found in Refs.[74-75].\\
\indent
With this background, we consider the force constant ($k=1$) for all sites are equal. We take $M_{1}=0.2$ and $M_{l}=M_{1}+0.2\times(l-1)$ with $l=2, 3, ..., N$. Figure \ref{fig7} displays the plot of rectification ratio with chain size for mass graded system with different environmental spectrum.
While at high temperatures
and for $T << T_a$ the effect is negligible and $J_+\sim  J_-$. On the other hand, $J_+$ and $J_-$ deviates significantly  at low temperatures and for large bias, with the current being larger in the direction of
increasing masses. It is also seen that the rectification ratio increases with chain size, an observation that can be reasoned by the growing mass ratio along the chain. In order to better understand the mechanism of thermal rectification one can inspect the temperature profile for the rectifying system. One may observe a reflection symmetry of the temperature profile with respect to the average temperature at high temperatures, when the temperature bias is reversed [51]. Further we observe an asymmetry in the temperature profile in the deep quantum domain beyond linear response regime. The
temperature gradient at the chain center is larger when the
light masses are in contact with the hot bath than the gradient
generated in the reversed case [51]. Furthermore, the asymmetry in the temperature profile of the Ohmic bath is larger compare to that of the sub-Ohmic and super-Ohmic cases. This is reflected in the rectification plot (Figure \ref{fig7}) which shows largest rectification ratio for the Ohmic bath.
\section{Heat transport across molecular junction}
In this section, we analyze the behavior of the heat current
as a function of molecular length, temperature, and
phonon-phonon scattering rate ($\gamma_I$). The main objective is to demonstrate that the
QSCR method can reproduce qualitative behaviour of the heat transport across anharmonic molecular junctions (MJs).
Although, the first-principle simulation results of quantum thermal
conduction in anharmonic MJs are unavailable, but examination on heat transport
in nanostructures [8-10], such as self-assembled monolayers,
amorphous polymers, and nanowires, exhibit some typical
features, like: (i) ballistic to diffusive transition
dynamics and the recovery of the Fourier’s law of thermal
conduction with increasing length and (ii) the heat current as a function of temperature shows a turnover [6,53].\\
\indent
Further we make a comparative study on the heat transport across MJs based on the following three methods:
(i) Using QSCR method we numerically solve Eqs. (\ref{eq:9}) and (\ref{eq:10}) and this nonlinear method enables us to catch far from equilibrium behaviour (large $\Delta T$), effective anharmonicity (large $\gamma_I$), and quantum effects ($\hbar\omega_0/T_a > 1$) which is introduced through the Bose-Einstein statistics, (ii) In the small temperature gradient, one can use Quantum linear-response approximation with the help of Eq.\,(\ref{eq:11}), (iii) while using Eq.\,(\ref{eq:12}) one can obtain Classical limit results with the help of
$T_a > \hbar\omega_0$ .\\
\indent
Furthermore, we compare the results for each method for the  three different bath spectrum: (i) Ohmic, (ii) Sub-Ohmic, (iii) Super-Ohmic environments.
\subsection{Length dependence}
\begin{figure}[b]
  \centering
  \includegraphics[width=\columnwidth]{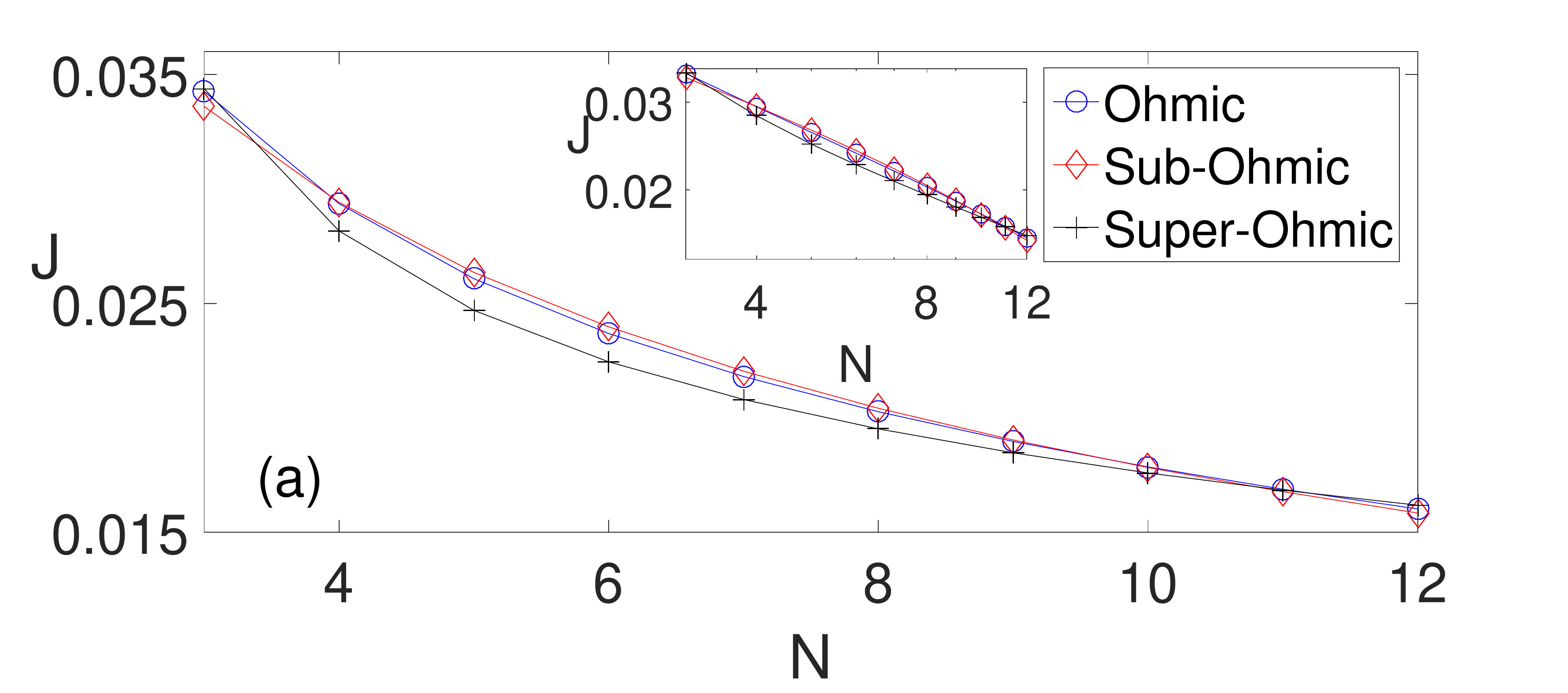}
  \includegraphics[width=\columnwidth]{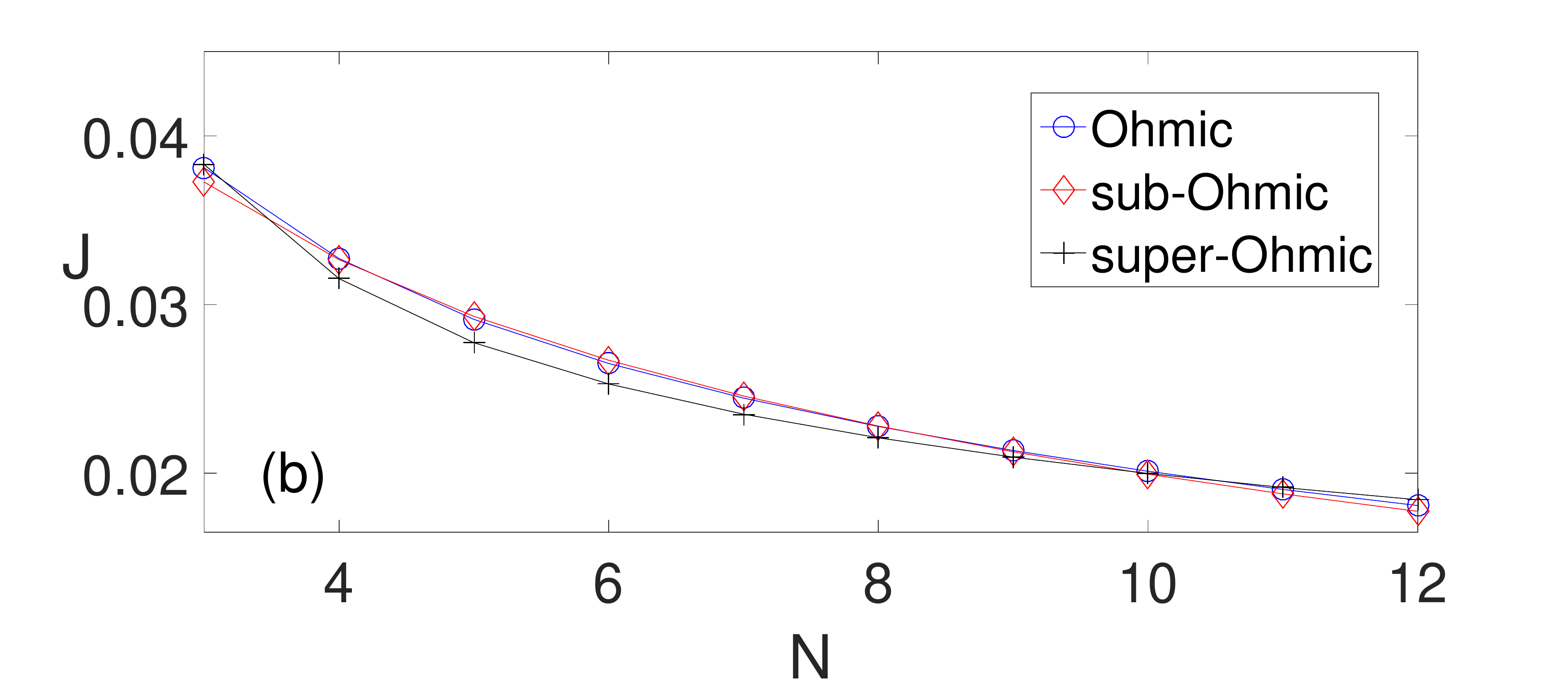}
  \includegraphics[width=\columnwidth]{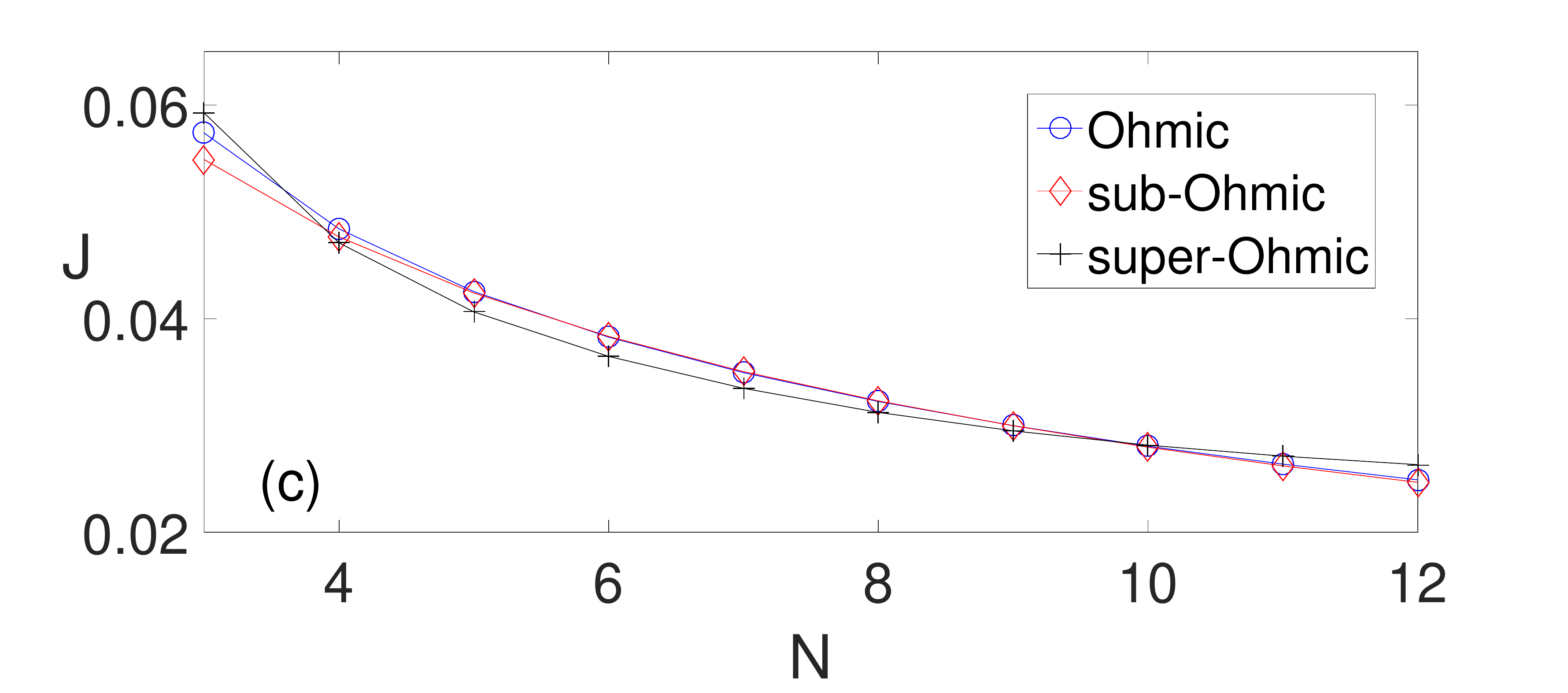}
  \caption{Heat current as a function of chain size using the (a) QSCRM, (b) quantum linear-response, and (c) classical calculations for Ohmic (o), sub-Ohmic ($\diamondsuit$), and super-Ohmic (+) environments. Parameters are $\beta_{1}=1$, $\beta_{N}=5$, $\gamma_{L}=\gamma_{R}=0.2$, $\gamma_{I}=0.8$. Here $\beta\equiv1/T$ is the inverse temperature. The temperature difference $\Delta T=T_{1}-T_{N}=0.8$.
  The average temperature $T_{a}=(T_{1}+T_{N})/2=0.6$. The inset in (a) displays the log-log scale plot.}\label{fig8}
\end{figure}
\begin{figure}[t]
  \centering
  \includegraphics[width=\columnwidth]{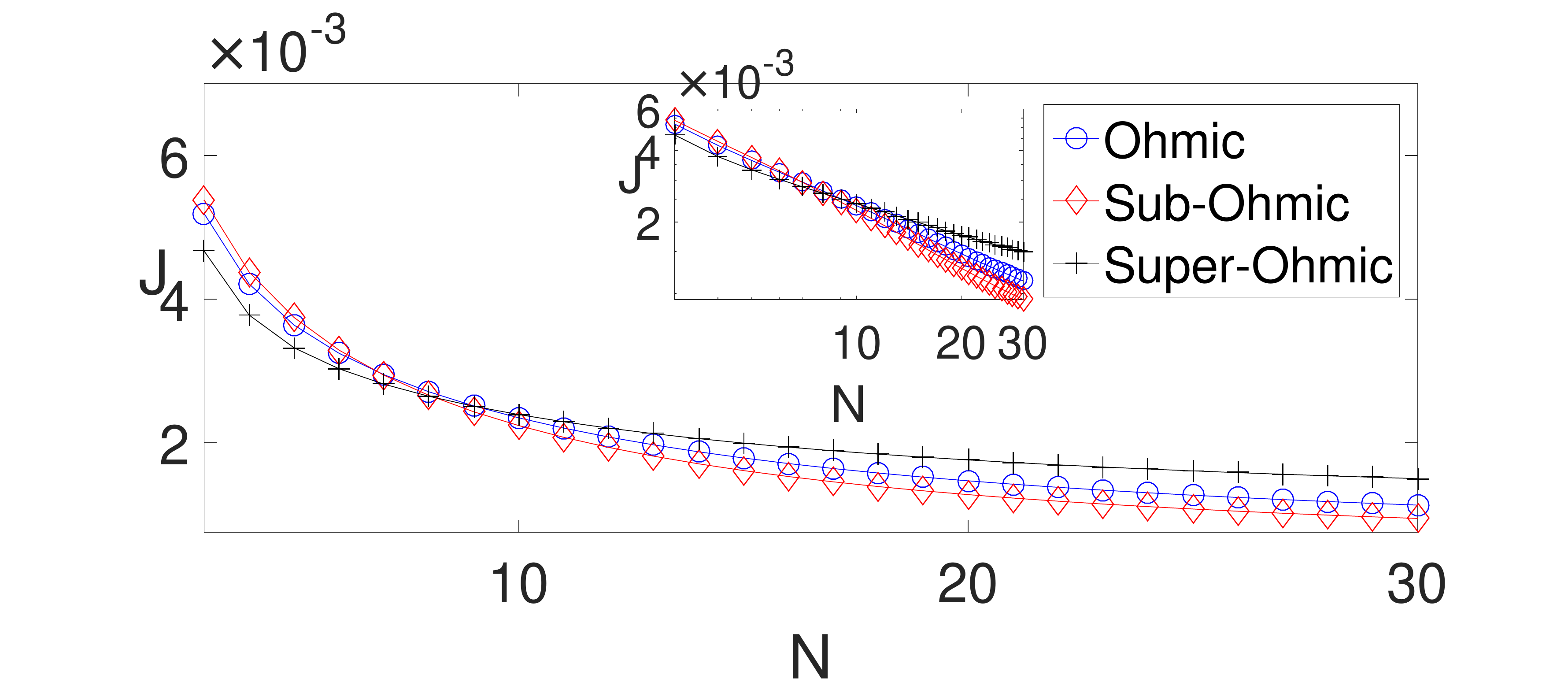}
  \caption{Heat current as a function of chain size using the QSCR method for Ohmic (o), sub-Ohmic ($\diamondsuit$), and super-Ohmic (+) environments. Parameters are $\beta_{1}=2$, $\beta_{N}=4$, $\gamma_{L}=\gamma_{R}=0.2$, $\gamma_{I}=2$. Here $\beta\equiv1/T$ is the inverse temperature. The temperature difference $\Delta T=T_{1}-T_{N}=0.25$. The average temperature $T_{a}=(T_{1}+T_{N})/2=0.375$. The inset displays the log-log scale plot.}\label{fig9}
\end{figure}

In this subsection, we observe the behavior of the heat current with chain length which is
displayed in Fig.\,\ref{fig8} for short chain and Fig.\,\ref{fig9} for long chain. The conductance of quantum harmonic systems for short molecules is usually make a non-monotonous behaviour with length [6, 28–30]. Furthermore, the heat current behaviour of such short molecules depend on the normal mode spectrum,
temperature, the nature of the contacts, and the properties
of the solids also dictate the heat current behaviour of such short molecules. Considering a 1-D ordered chain one can observe a ballistic behavior for harmonic systems. The QSCR method captures the anharmonic effect properly and it causes the decay of the current with molecular length, reflecting an increase in the thermal
resistance. On the other hand, classical simulations always overestimate the current. In the quantum linear-response regime, the heat current is qualitatively close to the QSCR
calculations.\\
\indent
Furthermore, one may find whether the SC reservoirs can reproduce
Fourier’s law of thermal conduction, i.e., $J\sim \frac{\Delta T}{N}$ [25]. One can observe that
 the temperature drop along the chain is relatively large for short molecules. As one increases the chain length one find that (inset) heat current behaviour approaches the Fourier’s law $J\sim N^{-1}$. The QSCR method is computationally costly for large N. We perform the simulations for $N=30$ to demonstrate the heat current behavior for long chain (see Fig.\,\ref{fig9}).\\
 \indent
 The behavior of heat current with chain length is displayed in figure \ref{fig8} for different environmental spectrums. For short molecule the temperature drop along the chain is large. Here we examine whether the SC reservoirs can satisfy the Fourier's law of thermal conduction, i.e., $J\varpropto\frac{\triangle T}{N}$ [25].
  For $N=12$, a log-log plot of conductance length (see inset in Fig.\,\ref{fig8}(a)) reveals that the slope is same ( $\sim-0.54$) for both the Ohmic and sub-Ohmic cases whereas the slope is $\sim-0.53$ for the super-Ohmic environment. For $N=30$ the log-log plot (see inset in Fig.\,\ref{fig9}) reveals that the slopes are $\sim-0.66$, $\sim-0.77$, and $\sim-0.47$ for the Ohmic, sub-Ohmic and super-Ohmic environment, respectively. Thus, we can conclude that the heat current reaches Fourier's law for the sub-Ohmic case with a shorter chain length compare to the other two. On the other hand, super-Ohmic environment requires largest chain length to attain Fourier's law.
\subsection{Temperature dependence}
\begin{figure}[b!]
  \centering
  \includegraphics[width=\columnwidth]{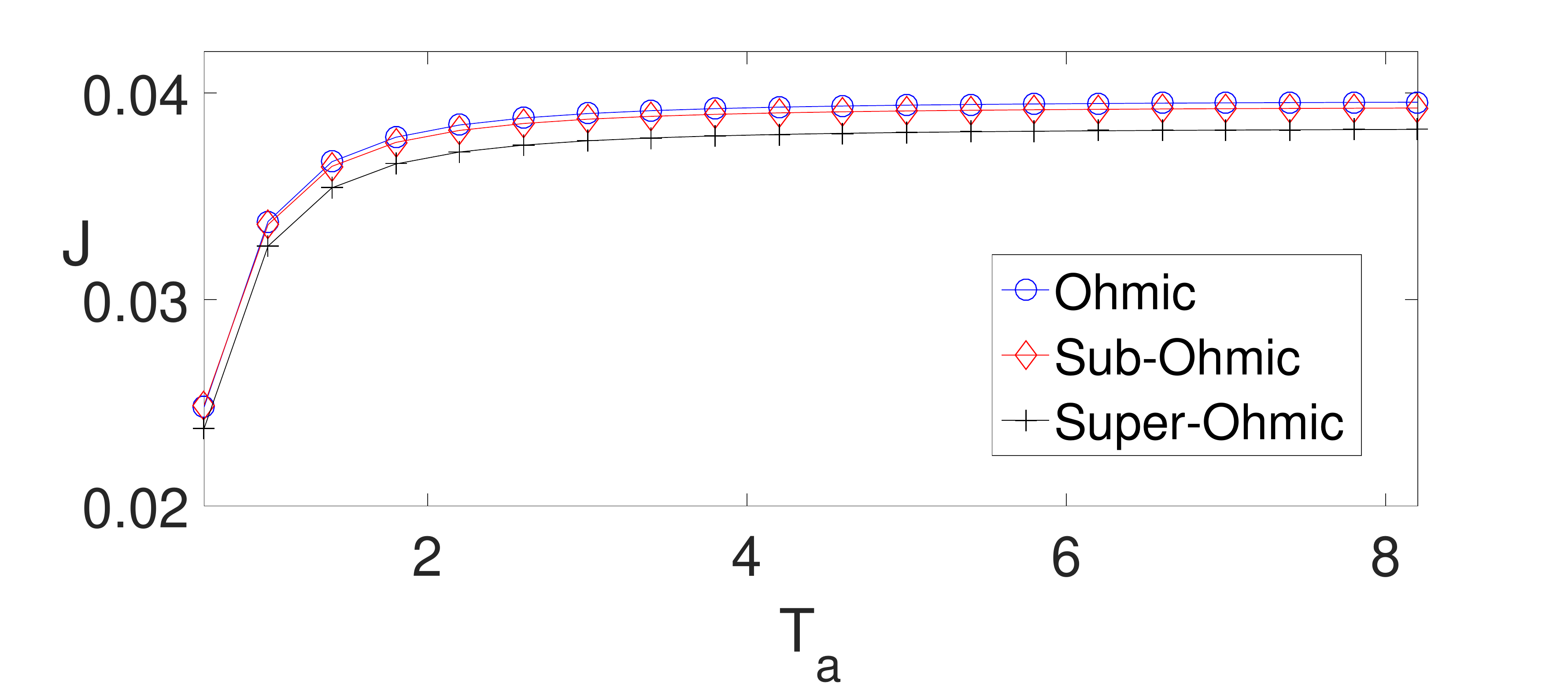}
  \caption{Heat current as a function of average temperature using the QSCR method for Ohmic (o), sub-Ohmic ($\diamondsuit$), and super-Ohmic (+) environments. Here $N=10, \triangle T=0.8,\gamma_{L}=\gamma_{R}=0.2$. Temperature independent, $\gamma_{I}=0.4$, phonon scattering rates.}\label{fig10}
\end{figure}
\begin{figure}[t]
  \centering
  \includegraphics[width=\columnwidth]{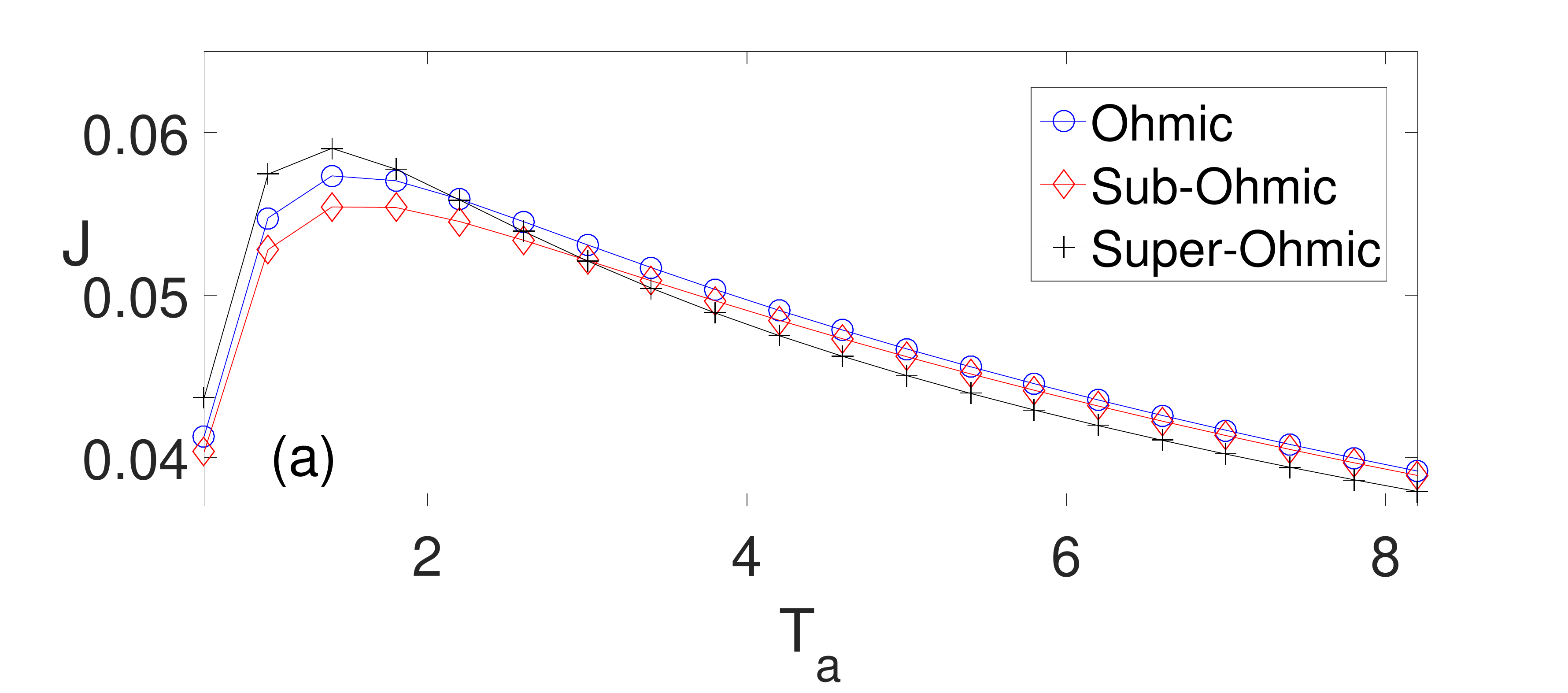}
  \includegraphics[width=\columnwidth]{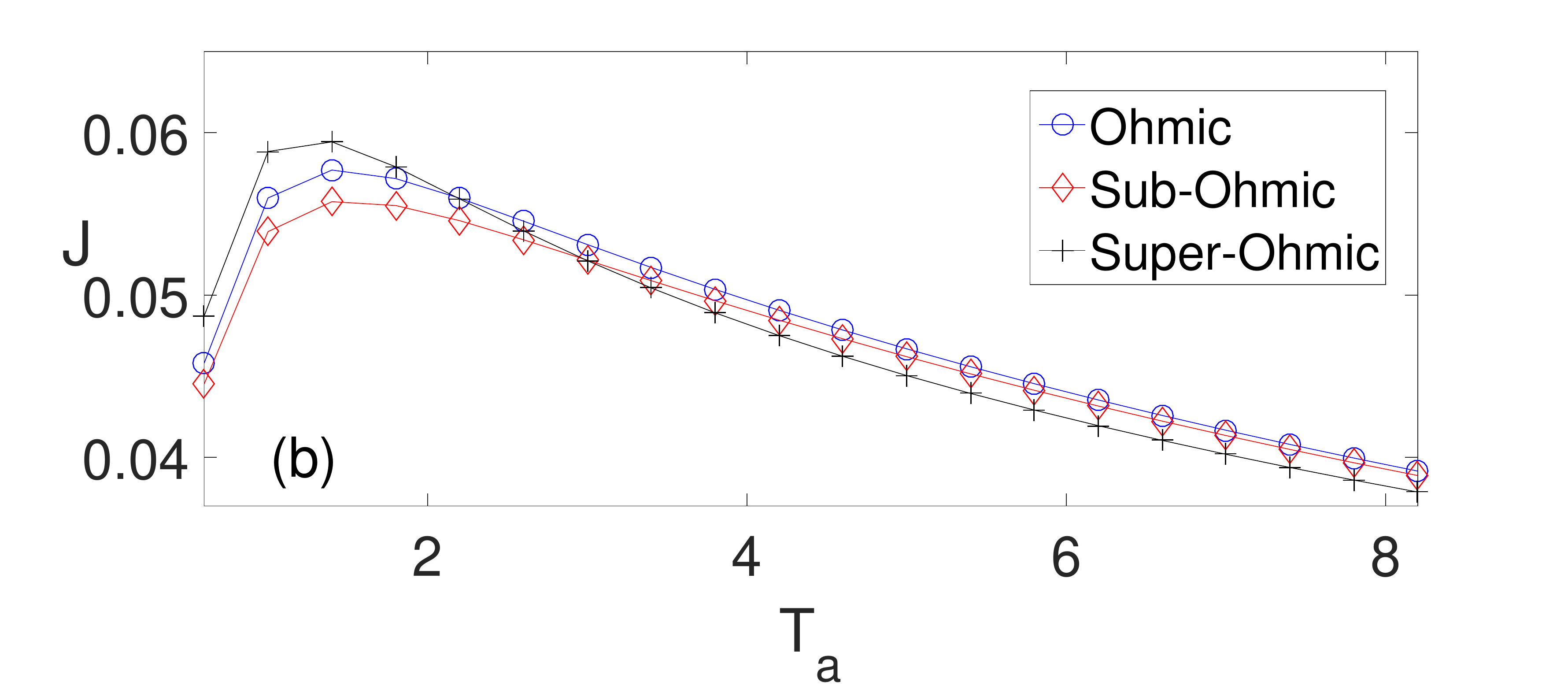}
  \includegraphics[width=\columnwidth]{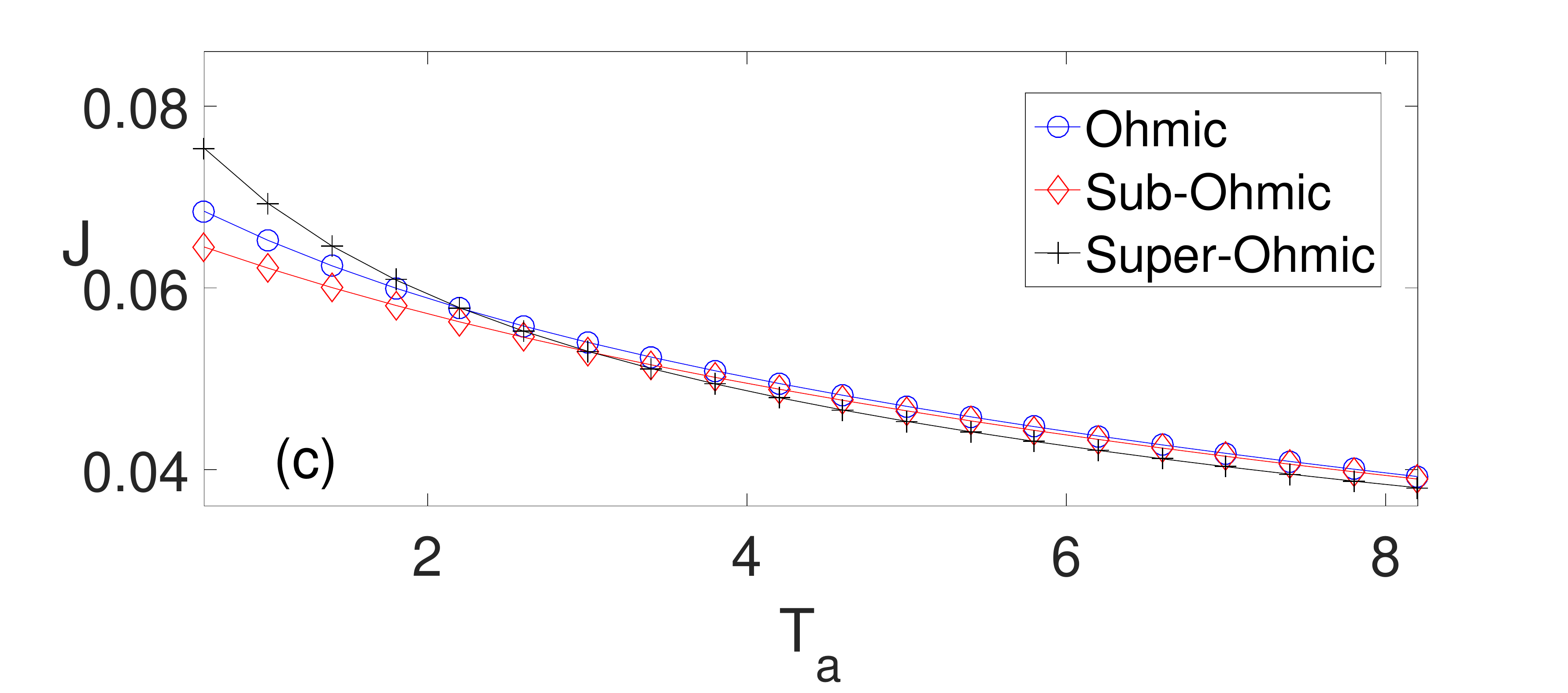}
  \caption{Heat current as a function of average temperature using the (a) QSCRM, (b) quantum linear-response, and (c) classical calculations for Ohmic (o), sub-Ohmic ($\diamondsuit$), and super-Ohmic (+) environments. Here $N=10, \triangle T=0.8,\gamma_{L}=\gamma_{R}=0.2$. Temperature dependent, $\gamma_{I}=0.05T_{a}$, phonon scattering rates.}\label{temp_gamma}
\end{figure}
We study the heat current as a function of temperature. The junction is biased following $T_{L,R} = T_a\pm \frac{\Delta T}{2}$. At low temperatures Classical results always exaggerate quantum simulations. On the other hand, QLR calculations may overestimate the QSCR results. Further, the simulation at high temperatures reveals that QSCR heat current saturates similar to harmonic predictions.The behavior of heat current with temperature is displayed in the Fig.\,\ref{fig10}. We see that at high temperature the results saturate for all the three environments. On the other hand, bulk [7] and nano-structure materials [70] typically show a turnover behavior
with temperature. The thermal conductance initially increases
with $T_a$ and it decays with $T_a$ as one goes beyond $T_a\simeq \hbar\omega_0$ due to increasing significance of the (inelastic) Umklapp phonon-phonon
scattering process. Hence, we can expect a significant role of anharmonic interactions in MJs as the temperature is
increased. As we increase the temperature more number of modes become active and more pathways open up for the redistribution of vibrational energy which results in the increase of thermal resistance.\\
\indent
One can capture the turnover behavior in the QSCR
method by making the internal scattering rates as a function of
temperature, $\gamma_I(T_a)\propto T_a^p$,
 $p > 0$. This results in the increase of
anharmonic effects as we increase the temperature, eventually suppress the current. Overall, the increase of $\gamma_I$ with temperature leads to a turnover behavior. The turnover behavior for (a) QSCRM, (b) quantum-linear response and (c) classical calculations for three different bath spectrum is shown in the Fig.\,\ref{temp_gamma}. We see that the heat current increases for low temperature upto certain values of $T_a$ and then decays for high temperature limit in QSCRM and quantum-linear response case. However, the heat current decays for all temperature limits in classical calculation. When we consider temperature dependence of scattering rate, the heat current initially shoots up much more for the super-Ohmic environment than that of Ohmic and sub-Ohmic cases. This is quite different than that of temperature independent scattering rate (see Fig.\,\ref{fig10}). Above certain values of $T_a$, the heat current behaviour follows the usual decreasing nature where the decreasing rate is maximum for the super-Ohmic environment. Although, classical calculation is incapable to demonstrate such behaviour.\\
\indent
In the present study we consider $\gamma_I\propto T_a$ (with p=1) in consistency with the study of Refs.[16,53].
One can actually show  that different scaling for the heat current with $T_a$ can be achieved
by adjusting the temperature dependence of the scattering rate. We examine a linear dependence. Further a study on quadratic form can be found in Ref.[53]. Although the later option is more physically imitates the
phonon-phonon collision processes, as the occupation factor
of each incoming phonon scales with temperature [53]. One can demonstrate a turnover behavior of the current
with temperature around the same point for the both models, but they show distinct
scaling laws. Therefore, by modifying the functional form $\gamma_I(T_a)$ ( physically represents to different inter-molecular-vibrational rate) we can emulate different
types of scattering processes.
\subsection{Scattering rate dependence}
\begin{figure}[b!]
  \centering
  \includegraphics[width=\columnwidth]{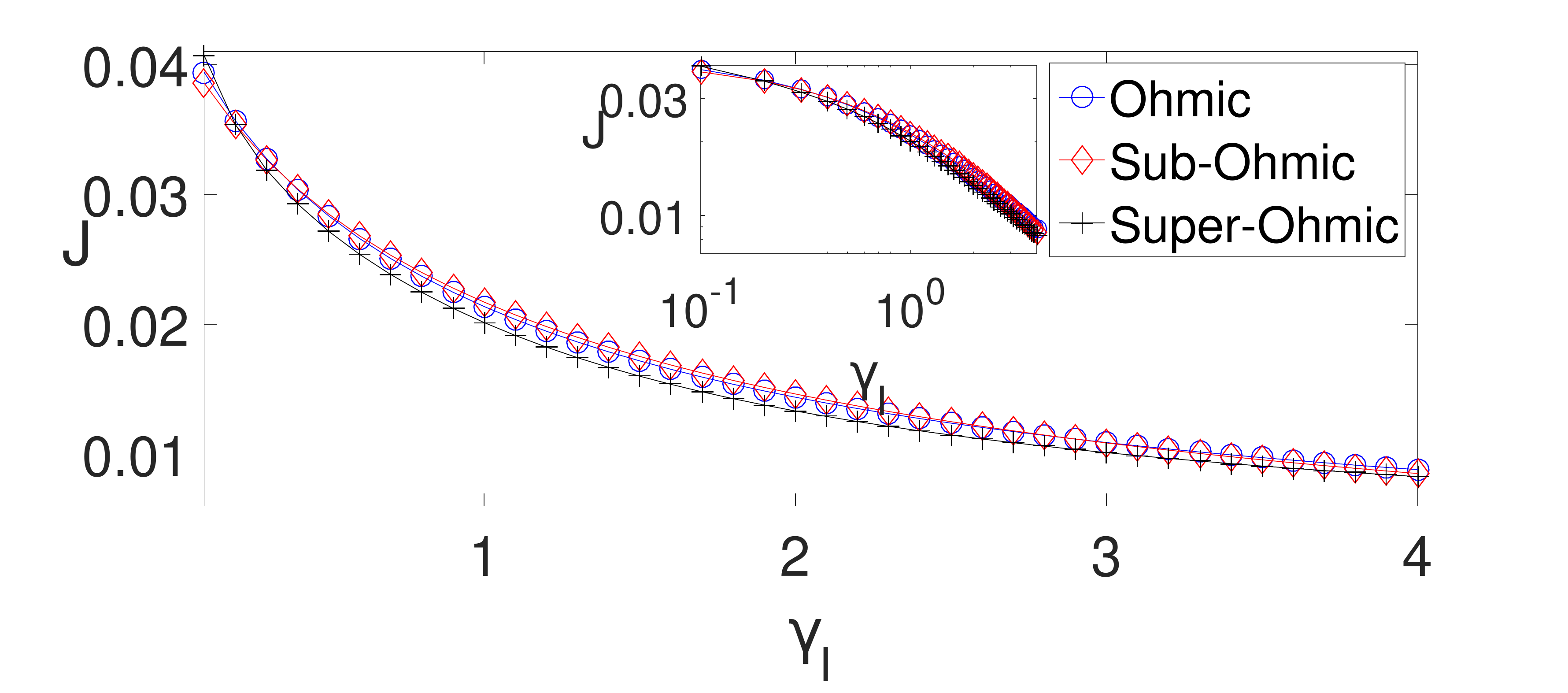}
  \caption{Heat current as a function of phonon-phonon scattering rate $\gamma_{I}$ using the QSCR method for Ohmic (o), sub-Ohmic ($\diamondsuit$), and super-Ohmic (+) environments. Here we take $N=6,\beta_{1}=1,\beta_{N}=5,\gamma_{L}=\gamma_{R}=0.2$. Here $\beta\equiv1/T$ is the inverse temperature. The temperature difference $\Delta T=T_{1}-T_{N}=0.8$.
  The average temperature $T_{a}=(T_{1}+T_{N})/2=0.6$.}\label{fig11}
\end{figure}
It is well known that as we increase the scattering rates of the SC baths $\gamma_I$, it also increases the amount
of anharmonicity in the model. The increase of $\gamma_I$ ultimately enhances the phonon splitting and recombination
processes. As a matter of fact the excitations are scattered between the SC baths, thus the thermal resistance
grows with the increase of $\gamma_I$. Similar to earlier cases one may observe that the heat current is escalated in the classical regime, and it is slightly overvalued in the
QLR method. It is known from Refs. [51,53] that the thermal conductance is inversely proportional to the phonon scattering rate. But, the short molecules display deviation from this trend and it indicates that the
current is dominated by the direct (harmonic) left-to-right transmission contribution. But, as we increase the chain length it
approaches the theoretical limit. The effect of three different types of environment is quite similar in this perspective of scattering rate dependence (see Fig.\,\ref{fig11}).\\
\indent
Before concluding this section, we can summarize and explain all of our observed results of this section based on Kapitza resistance [76,77]. The interface thermal resistance (ITR) or the
Kapitza resistance, measures the interfacial resistance to heat
flow [76,77]. It is usually defined as $R=\frac{\Delta T}{J}$, where J is the heat flow (per unit area) and $\Delta T$ measures the temperature jump between two sides of the interface. The emergence of the Kapitza resistance can be thought of due to the heterogeneous vibrational properties
of the two different materials fabricating the interface at which the
energy carriers collide with each other. Usually the amount of relative
transmission of heat energy banks on the available energy states on each
of the two sides of the interface. It is observed that the ITR in our setup depends (a) on the
direction of the applied temperature bias and (b) the degree of overlap of the power spectra between two segments i.e. the left (L) and right (R) segments of the interface.  The underlying physical mechanism of the unsymmetrical ITR between two dissimilar segments at the interface can be connected to the
match or mismatch of the corresponding power spectra. If the overlap of power spectra between two sectors is large one can expect large amount of heat flow along that interface and vice versa.\\
\indent
 Let us first consider figure 12. As
temperature increases, the power spectrum of the super-Ohmic bath
shifts downward toward lower frequencies of the system and overlap or resonance of the spectrum between two sectors occur. In contrast, the power spectrum of the sub-Ohmic bath segment, however, shifts upward,
toward higher frequencies. Because of these opposite shifts, overlap between power spectrum decreases and thermal energy flow is lesser than that of the super-Ohmic case. The same reasoning is applicable for Figure 14. Considering Figure 15 one can explain the phenomena as follows : one can think that the scattering effect is less available for the sub-Ohmic spectra and the overlap of power spectra across the interface is largest for the sub-Ohmic case. On the other hand, one can expect large amount of scattering for the super-Ohmic bath and the overlap  is least for the super-Ohmic scenario.
\section{Phonon mismatch: environmental effect}
\begin{figure}[t]
  \centering
  \includegraphics[width=\columnwidth]{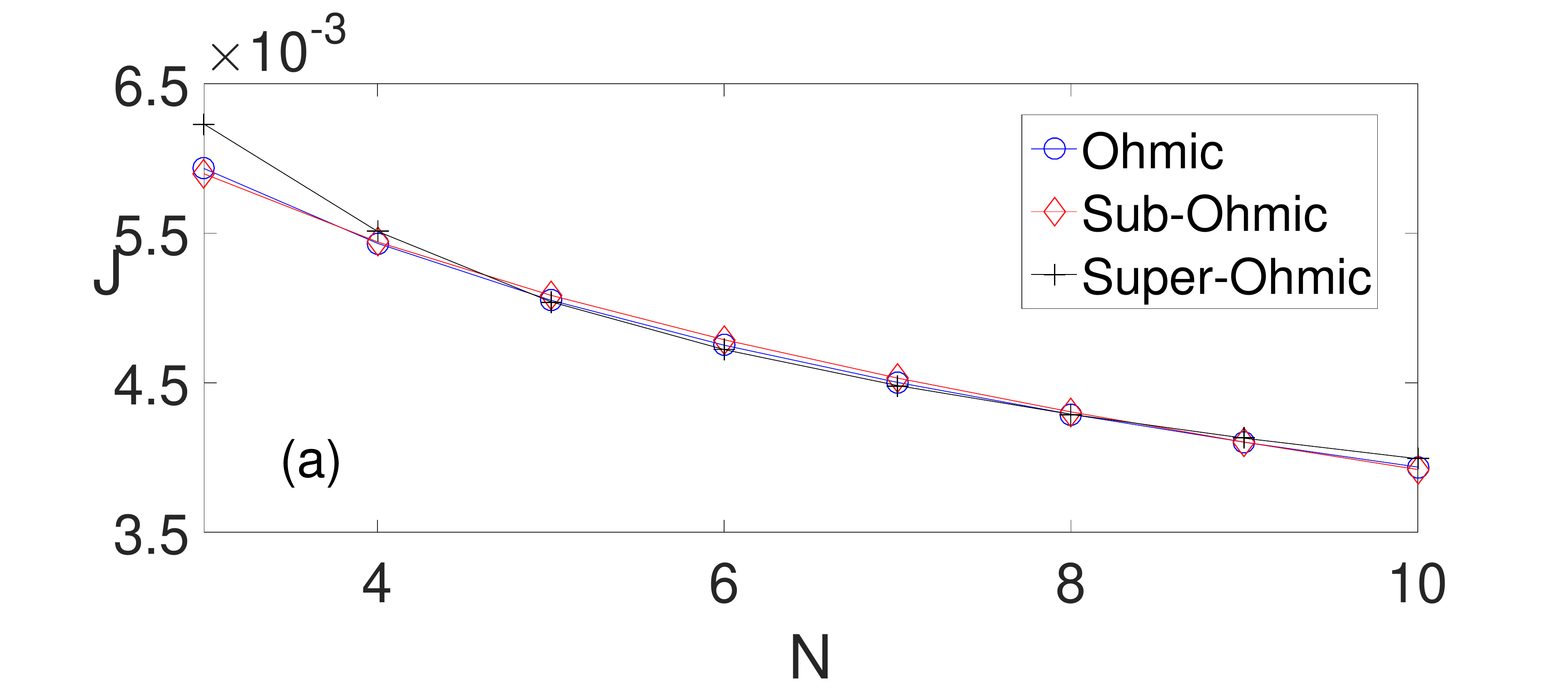}
  \includegraphics[width=\columnwidth]{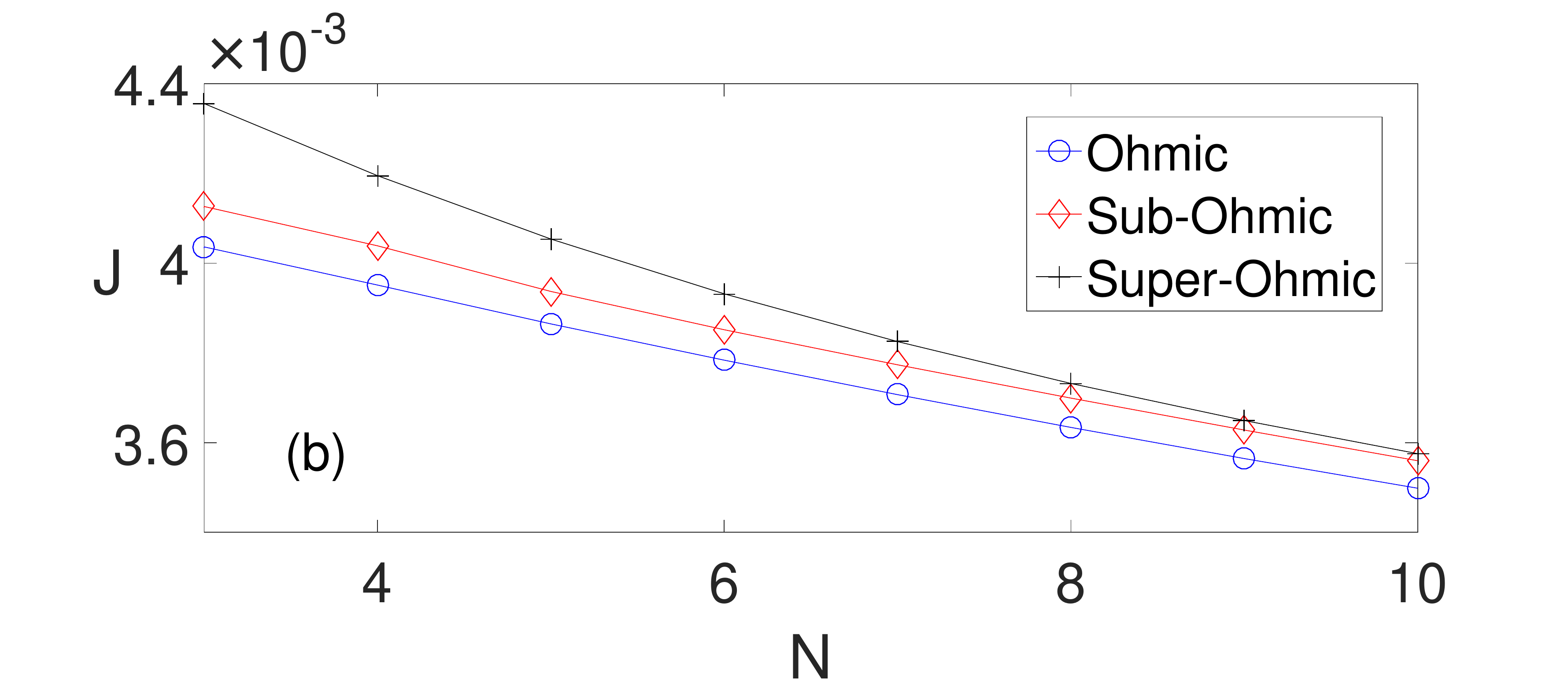}
  \caption{Heat current as a function of chain length with (a) Debye baths without mismatch with $\omega_{d,L}=\omega_{d,I}=\omega_{d,R}=4$, and (b) mismatched Debye baths with $\omega_{d,L}=4,\omega_{d,I}=1,\omega_{d,R}=1$. In both the cases we use the QSCR method for Ohmic (o), sub-Ohmic ($\diamondsuit$), and super-Ohmic (+) environments. We use $\beta_{1}=2,\beta_{N}=4,\gamma_{L}=\gamma_{R}=0.2,\gamma_{I}=0.8$. Here $\beta\equiv1/T$ is the inverse temperature, $\Delta T=0.25$, $T_{a}=0.375$.}\label{fig12}
\end{figure}

\begin{figure}[t]
  \centering
  \includegraphics[width=\columnwidth]{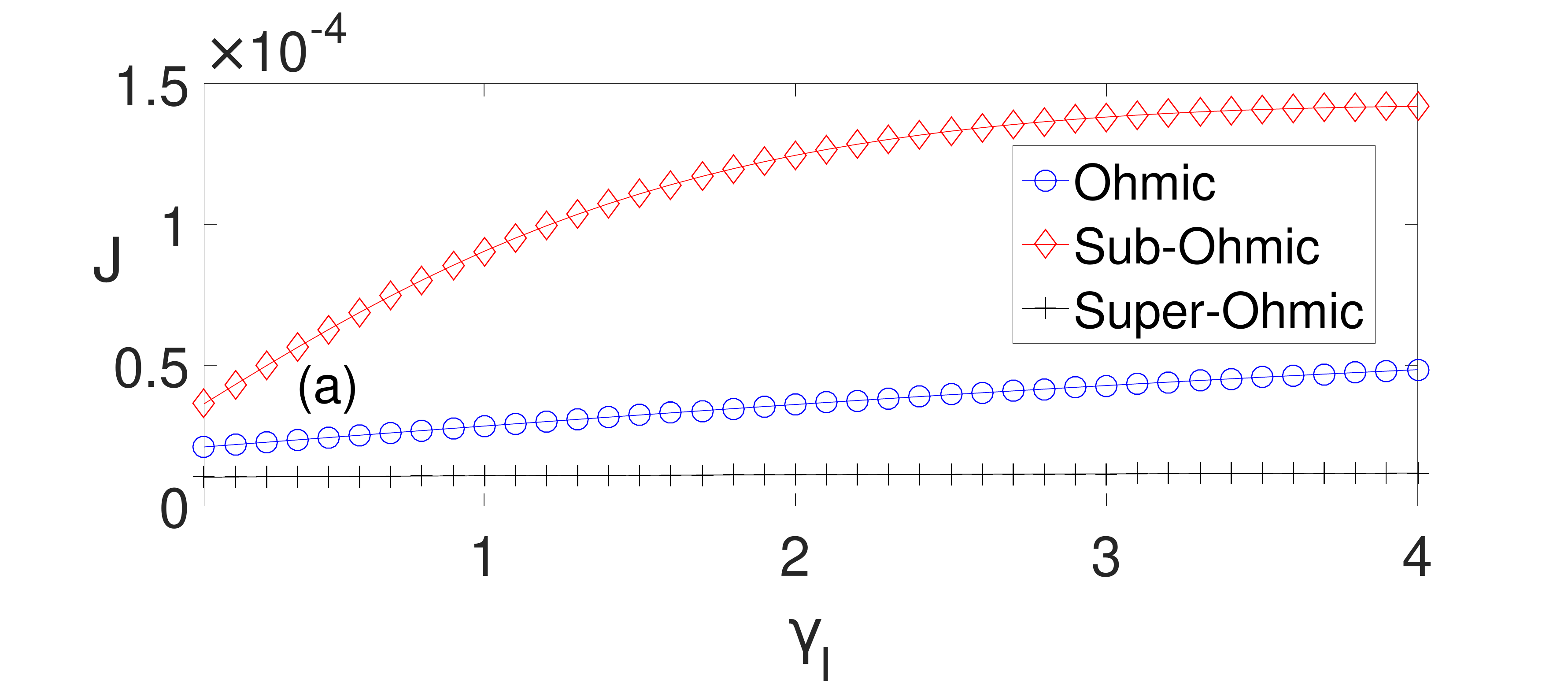}
  \includegraphics[width=\columnwidth]{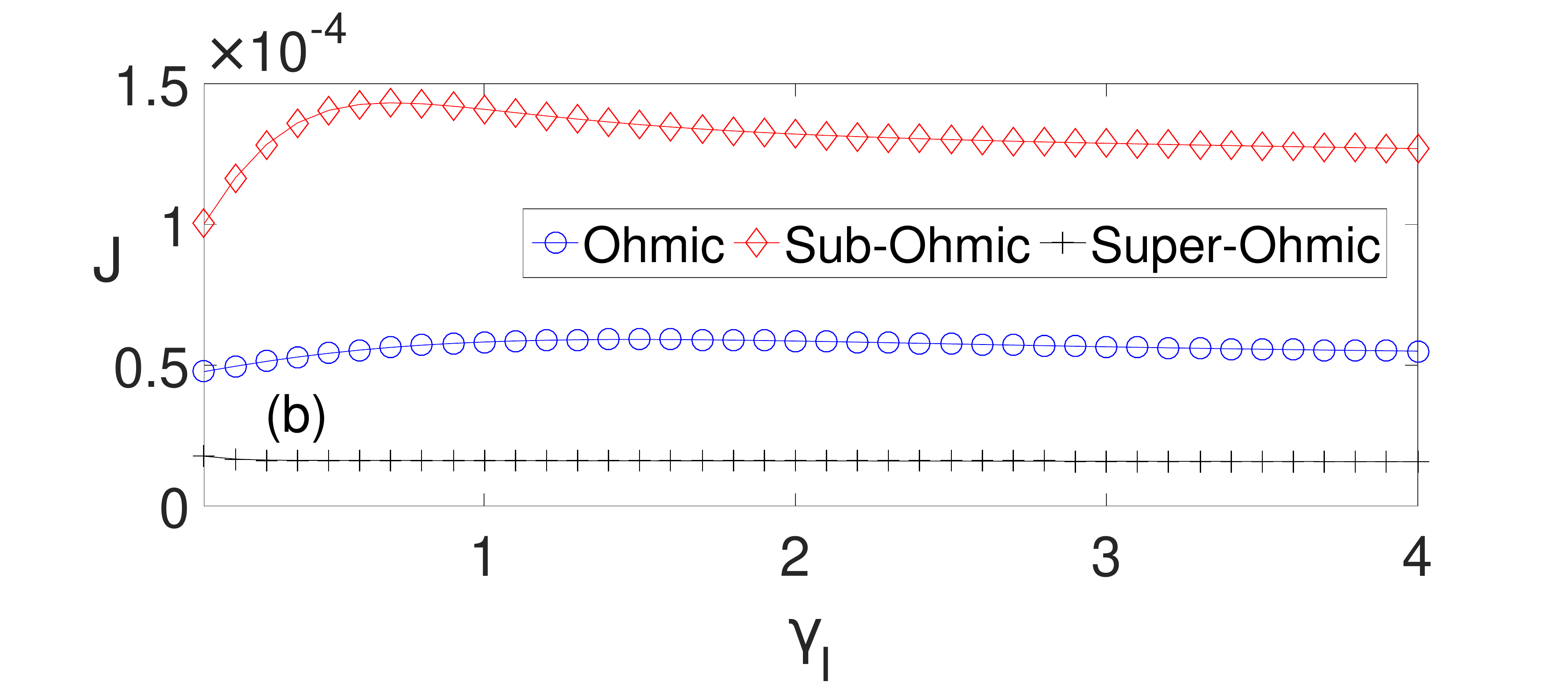}
  \caption{Heat current as a function of phonon scattering rate at low temperatures for (a) $N=5$ and (b) $N=10$. In both the cases we use the QSCR method for Ohmic (o), sub-Ohmic ($\diamondsuit$), and super-Ohmic (+) environments. The mismatch between the hot and cold bath is high, $\omega_{d,L}/\omega_{d,R}=10$. $\beta_{1}=2,\beta_{N}=4,\gamma_{L}=\gamma_{R}=0.2,\omega_{d,L}=1,\omega_{d,R}=0.1,\omega_{d,I}=0.1$. Here $\beta\equiv1/T$ is the inverse temperature, $\Delta T=0.25$, $T_{a}=0.375$.}\label{fig13}
\end{figure}

In the previous section we have shown that the QSCR method is able to capture
the anharmonic effects in quantum thermal transport. In the present section our focus is to
 study heat transport between two solid surfaces with structured and distinct phonon spectra. We use a simple  model in which the phonon spectra are distinguished by an exponential cutoff frequency $exp(-\omega/\omega_d)$, Eq.\,(\ref{eq:7}). Although we are using a simple set up but it is complex in nature as it includes the effect of anharmonicity, quantum effects, mismatched phonon spectra, and a far from equilibrium situation. The present section is devoted to study the effect of phonon mismatch between the two reservoirs on the heat current flow in MJs. The heat current usually drops as one increases the mismatch for a harmonic system but the inelastic scattering may help to tame it.\\
\indent
Here we are interested to study two cases: (a) exponential cut-off baths without mismatch, $\omega_{d,L}=\omega_{d,R}$, and (b) exponential cut-off baths with a vibrational mismatch, $\omega_{d,L}\neq\omega_{d,R}$. In Figure \ref{fig12}, we compare the result for the following three cases: (i) Ohmic, (ii) sub-Ohmic, and (iii) super-Ohmic environments. When we introduce vibrational mismatch, high frequency modes from left bath (hot) are barred to cross to the right cold bath unless and until energy is readjusted in the inner baths. But, the energy redistribution process is a slow method for short molecules and it results in suppression of heat current. Now, the dual role of SC reservoirs can be demonstrated. One can find that the increase of anharmonicity introduced by SC baths due to increase of phonon scattering rate $\gamma_{I}$ increases the thermal resistance and thus reduces the current as shown in Fig.\,\ref{fig11}. On the other hand, the increase of inelastic scattering rate ($\gamma_{I}$) can be helpful to overcome vibrational mismatch. In Fig.\,\ref{fig11}, we observe that as we increase $\gamma_I$ the heat current decreases. Further, if we consider Fig.\,\ref{fig13} the situation is quite different for the vibrational mismatch. For small $\gamma_I$ regime, as we increase the inelastic scattering rate the heat current increases. Thus the inelastic scattering is helping to overcome the vibrational mismatch in this regime. In this regime the anaharmonicity is helpful in transport in the MJs. However, after reaching certain value of $\gamma_I$ the SC baths increases the multiple scattering processes and dampen the heat current. The crossover between the two regimes depend on the length of the molecule (see Fig.\,\ref{fig13}). Further, one can observe that the effect is most prominent for the sub-Ohmic environment and least perceptible for the super-Ohmic case.\\
\indent
Next, we try to consider the experimental report in Ref. [58]. We consider molecular junctions with dissimilar phonon spectra at boundaries, captured by different Debye frequencies. We describe two such setups and study the heat current behavior with vibrational mismatch.
\subsection{Phonons up-conversion process}
\begin{figure}[t]
  \centering
  \includegraphics[width=\columnwidth]{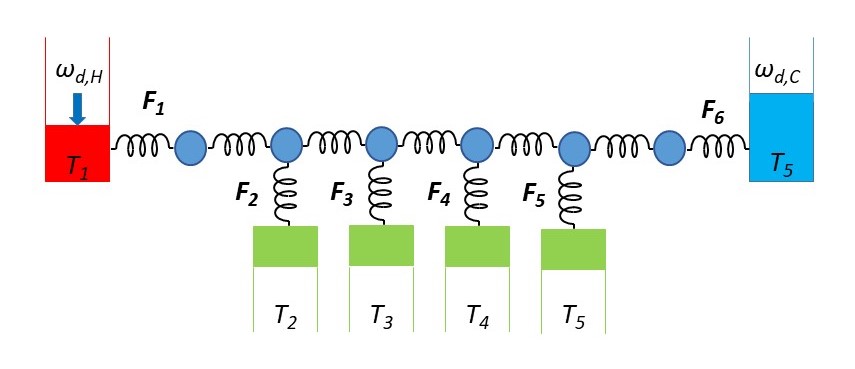}
  \caption{Scheme of the SC model of length $N=6$ with mismatched reservoirs for phonons up-conversion process.}\label{fig14}
\end{figure}
\begin{figure}[t]
  \centering
  \includegraphics[width=\columnwidth]{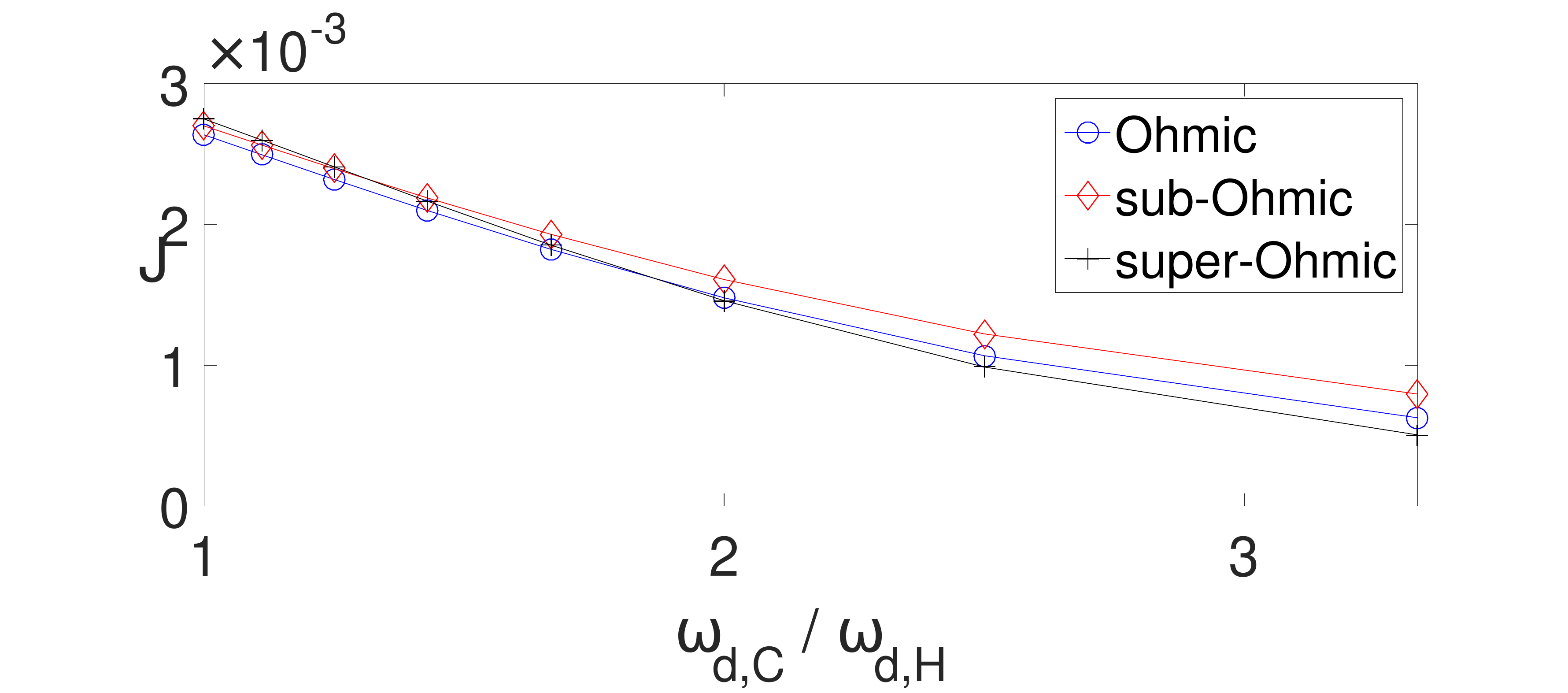}
  \caption{Heat current as a function of phonon mismatch with mismatched reservoirs, $\omega_{d,H}=1, 0.9,..., 0.3$, $\omega_{d,C}=1$, $\omega_{d,I}=2$ and $N=6, \beta_{1}=2,\beta_{N}=4,\gamma_{L}=\gamma_{R}=0.2,\gamma_{I}=0.8$. Here $\beta\equiv1/T$ is the inverse temperature, $\Delta T=0.25$, $T_{a}=0.375$. Here we the results of Ohmic (o), sub-Ohmic ($\diamondsuit$), and super-Ohmic (+) environments.}\label{fig15}
\end{figure}

The cold bath is fixed with a Debye frequency of $\omega_{d,C}=1$ as shown in Figure \ref{fig14}.
The Debye frequency of the hot bath is gradually reduced, starting from the same value as the cold bath. In this setup, the collisions of phonons  generate a high frequency mode (up-conversion) and it is helpful for transport. Figure \ref{fig15} demonstrates that heat current decays as one increases the vibrational mismatch (measured by the ratio of Debye frequencies) between the reservoirs. We study the results for Ohmic, sub-Ohmic, and super-Ohmic environments.
\subsection{Phonons down-conversion process}
\begin{figure}[t]
  \centering
  \includegraphics[width=\columnwidth]{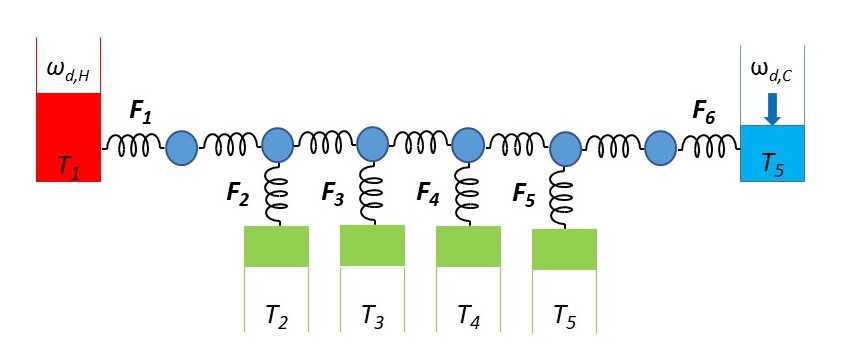}
  \caption{Scheme of the SC model of length $N=6$ with mismatched reservoirs for phonons down-conversion process.}\label{fig16}
\end{figure}
\begin{figure}[t!]
  \centering
  \includegraphics[width=\columnwidth]{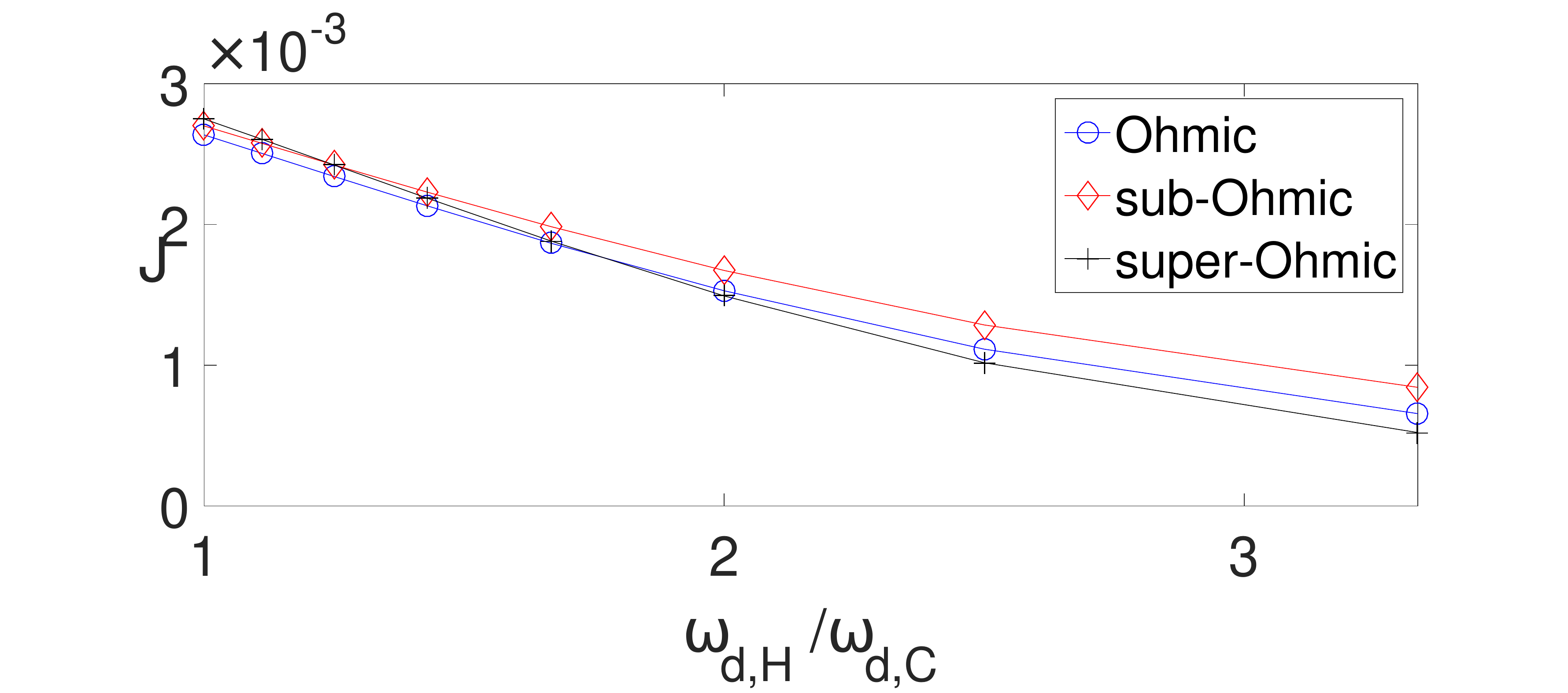}
  \caption{Heat current as a function of phonon mismatch with mismatched reservoirs, $\omega_{d,C}=1, 0.9,..., 0.3$, $\omega_{d,H}=1$, $\omega_{d,I}=2$ and $N=6, \beta_{1}=2,\beta_{N}=4, \gamma_{L}=\gamma_{R}=0.2,\gamma_{I}=0.8$. Here $\beta\equiv1/T$ is the inverse temperature, $\Delta T=0.25$, $T_{a}=0.375$. Here we the results of Ohmic (o), sub-Ohmic ($\diamondsuit$), and super-Ohmic (+) environments.}\label{fig17}
\end{figure}
In this setup, the hot solid is kept fixed and its Debye frequency is either make equal or higher than the cutoff frequency of the cold bath. As a matter of fact the high frequency modes can not cross the system in the harmonic limit. For this set up the phonons splitting of low frequency modes are useful for transport. One may find that
 the heat current decays as the vibrational mismatch
between the solids increases. One may further test the reaction of the cutoff frequency of the SC baths on the
heat current. If we choose SC baths at a low cutoff frequency,
their ability to scatter phonons reduces and as a result it approaches the
harmonic limit. On the other hand, when the SC baths cover the full spectra ($\omega_{d,I}$
large), the heat current reduces as the thermal resistance increases. In this process the SC baths basically dampen the heat current (see Fig.\,\ref{fig17}).
\begin{figure*}[t!]
  \centering
  \includegraphics[width=5cm]{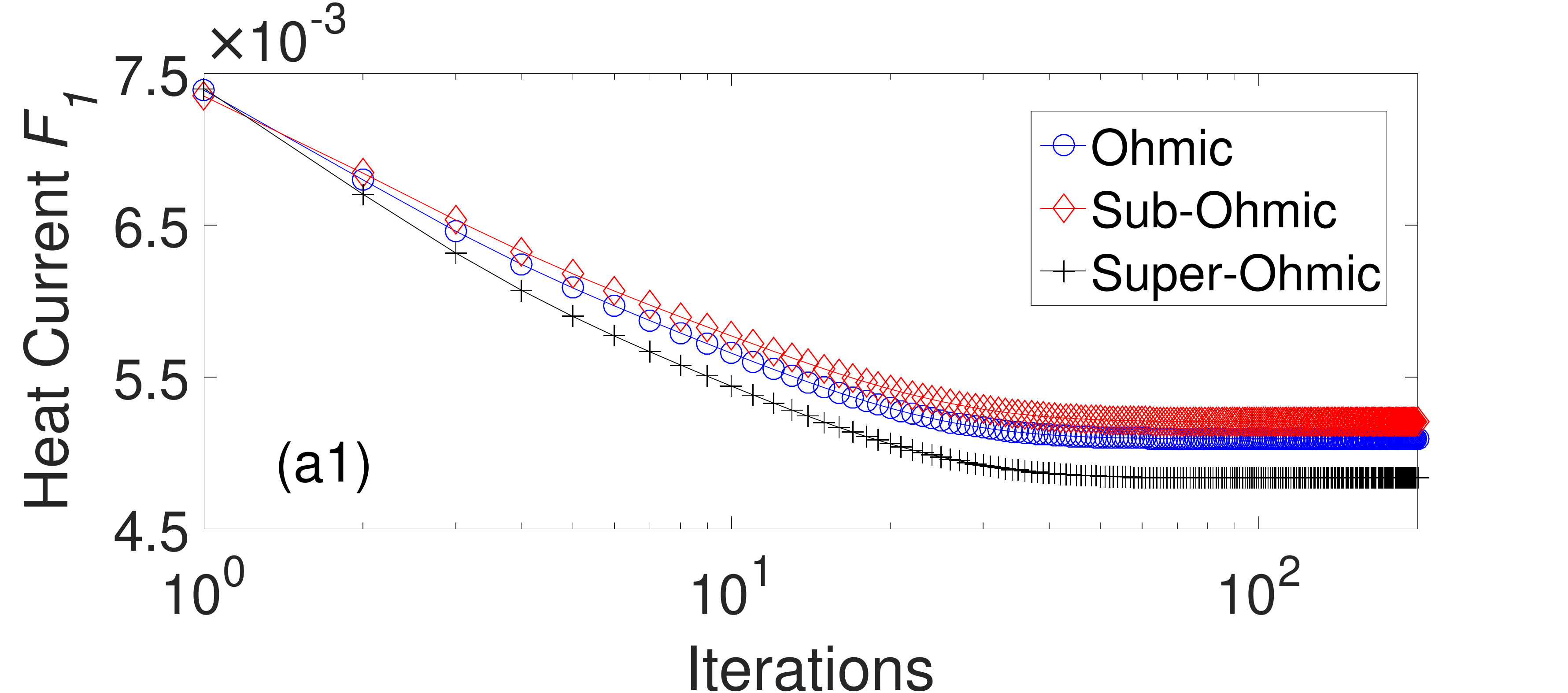}
  \includegraphics[width=5cm]{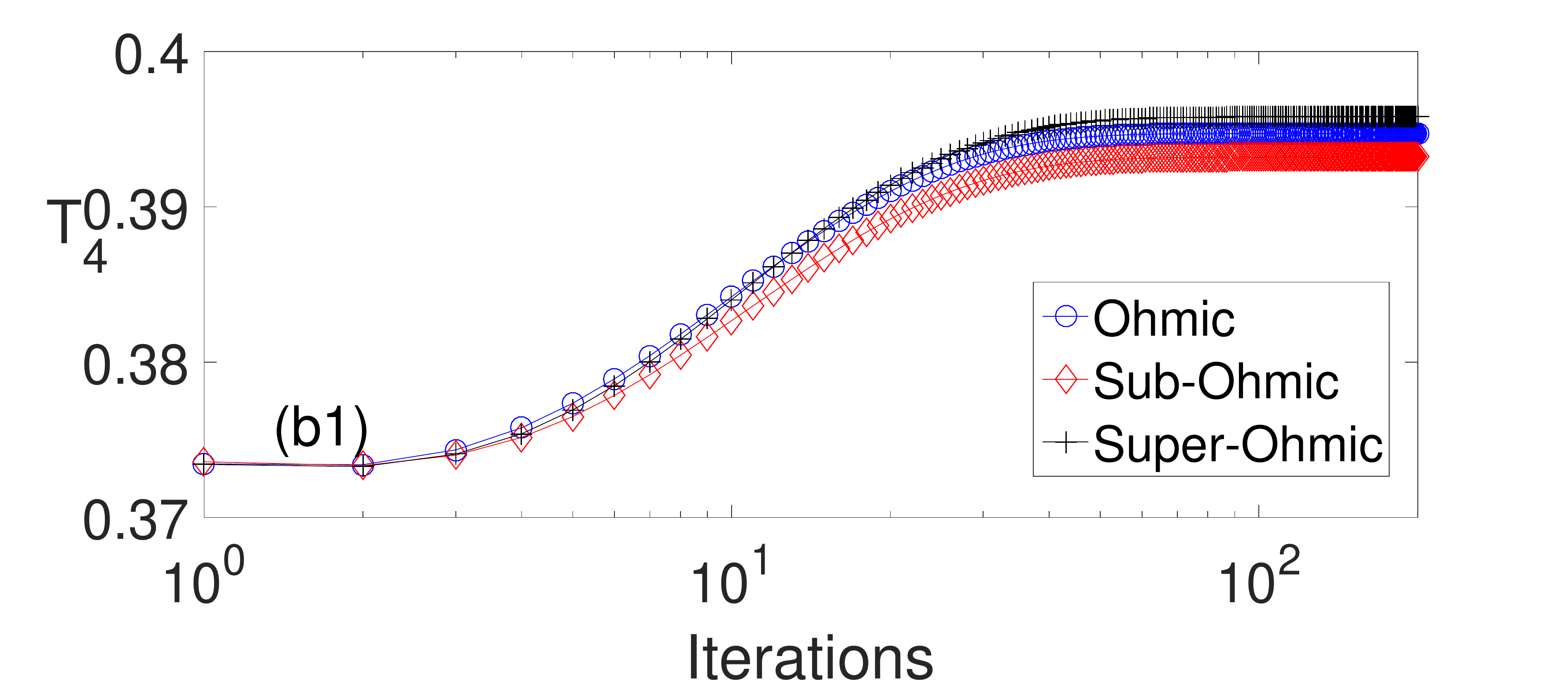}
  \includegraphics[width=5cm]{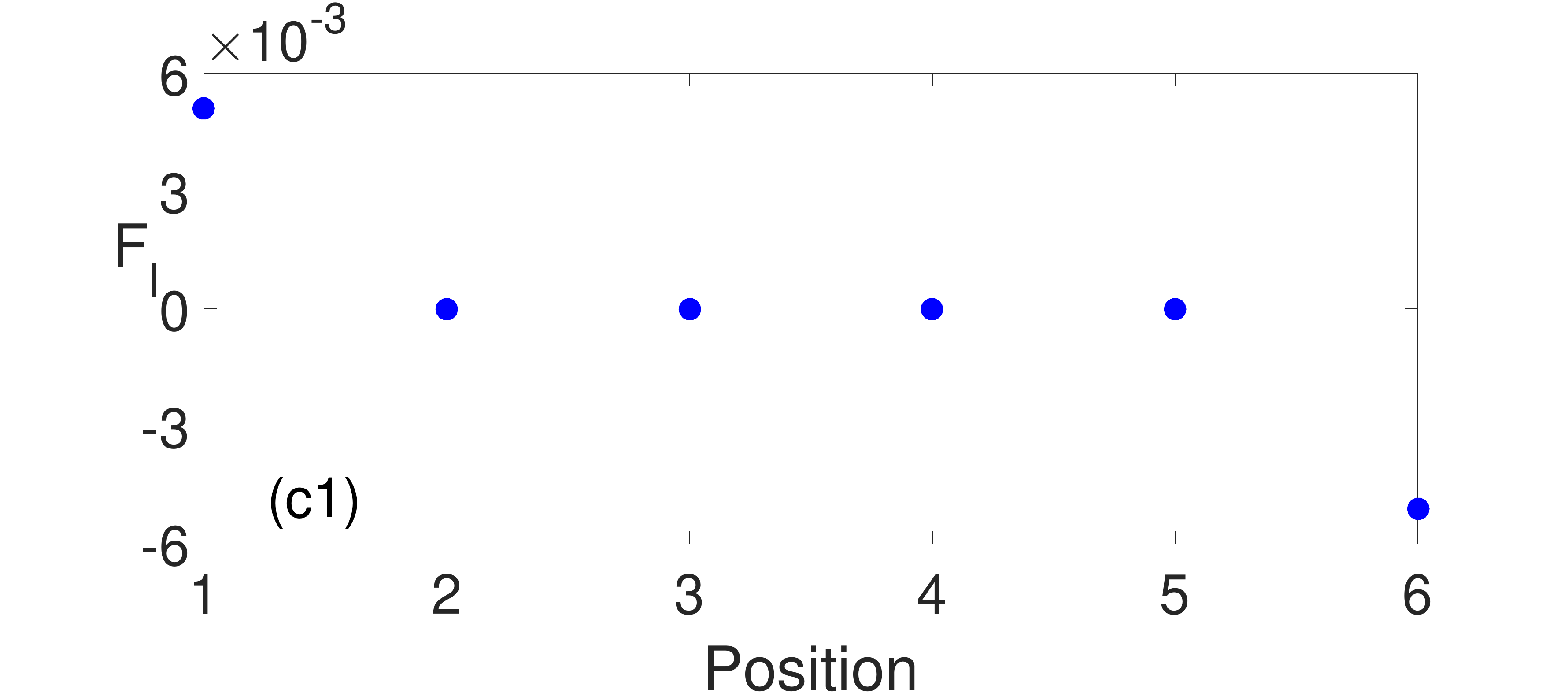}
  \includegraphics[width=5cm]{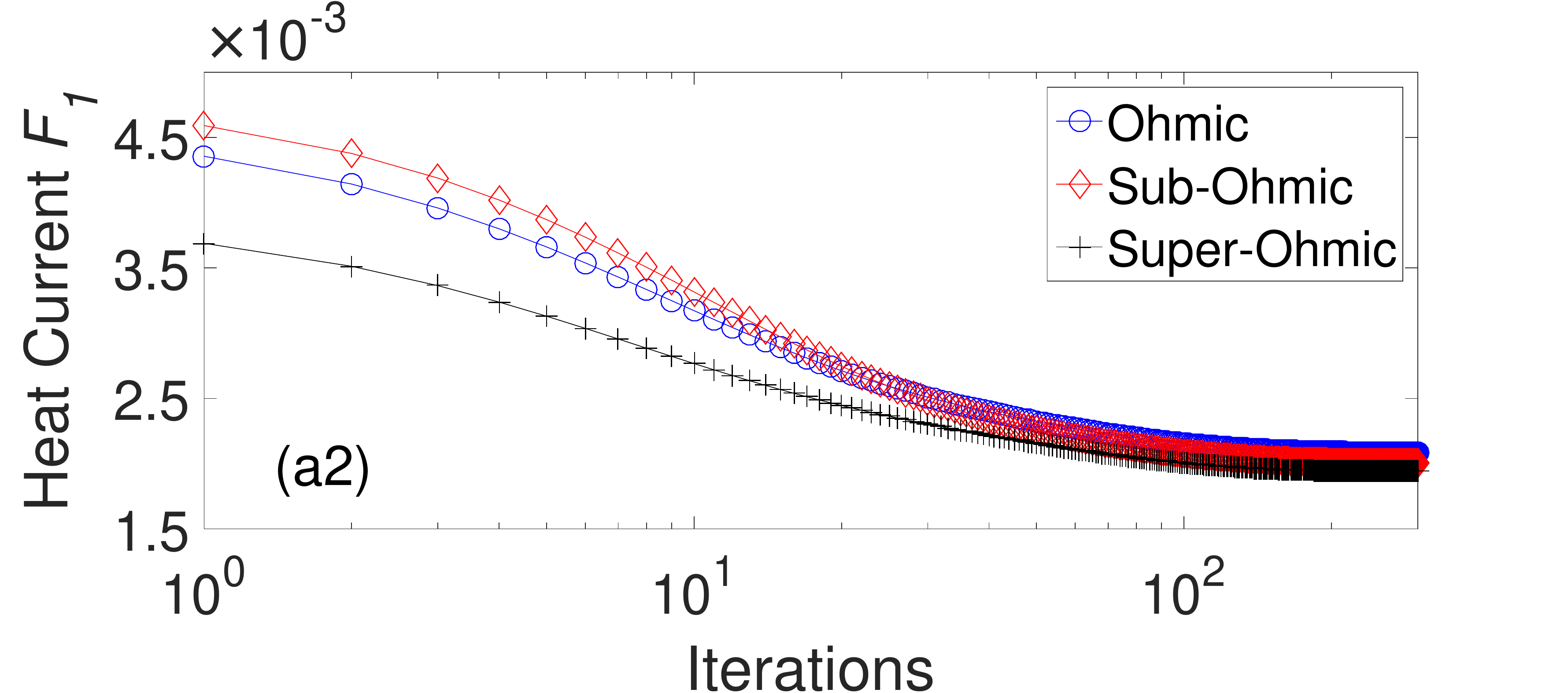}
  \includegraphics[width=5cm]{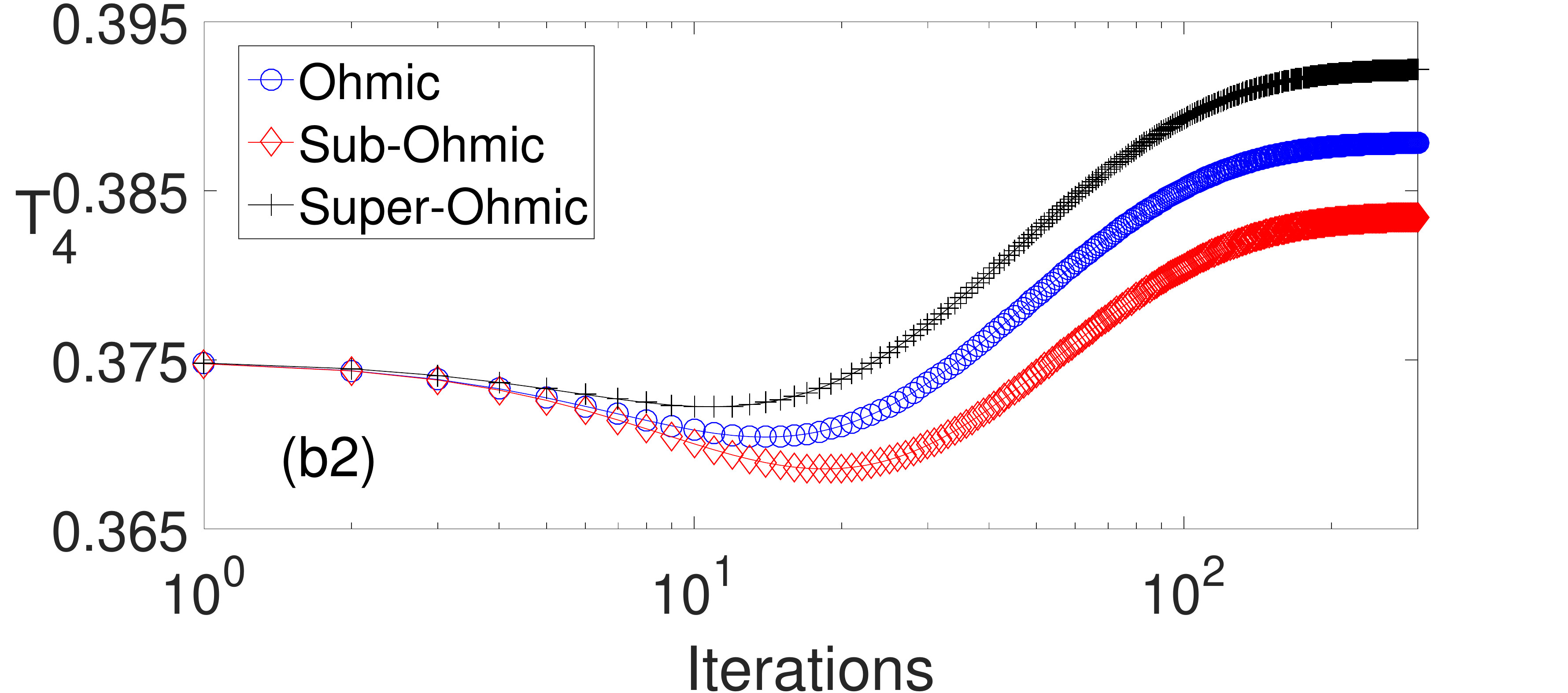}
  \includegraphics[width=5cm]{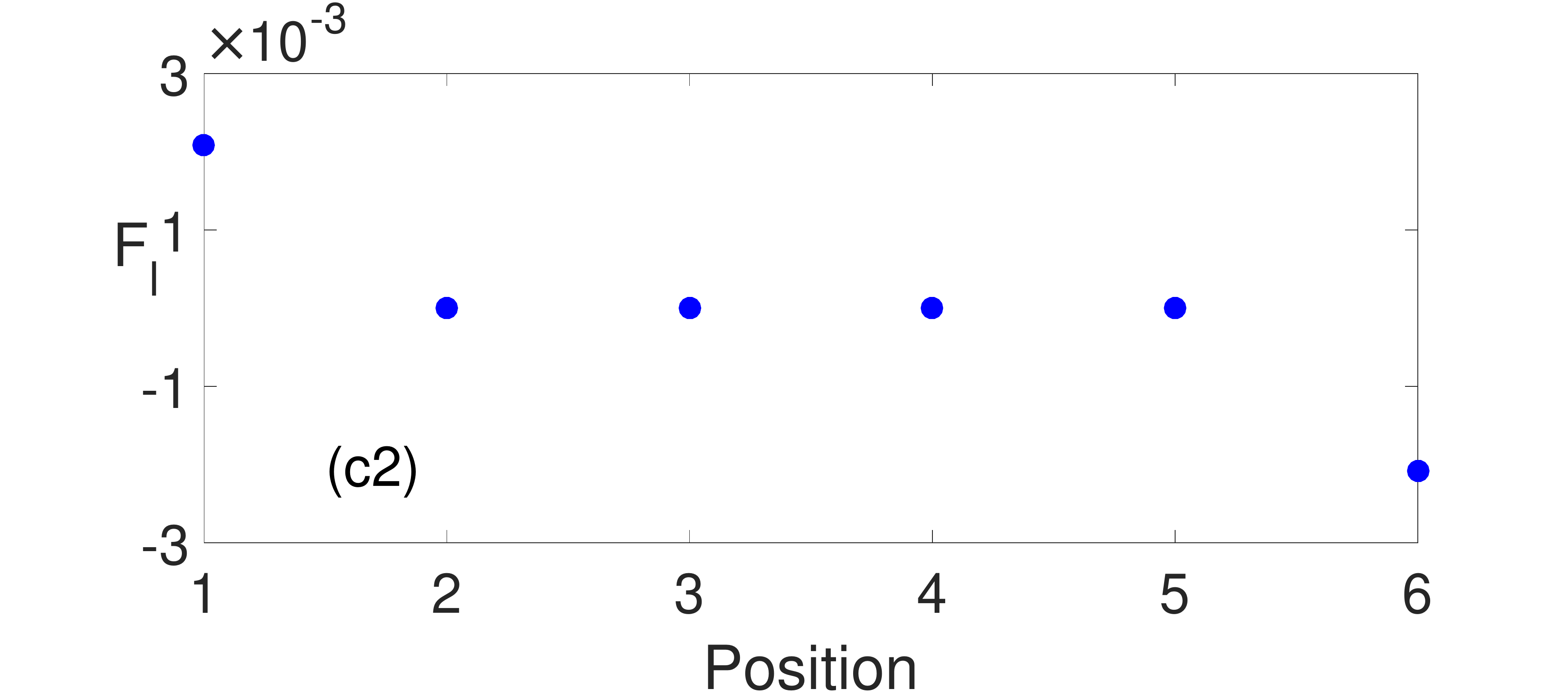}
  \caption{Convergence of the QSCR method with the increasing number of iterations using structured bath without exponential cut-off ($\omega_d\rightarrow \infty$) baths. Top ($\gamma_{I}=0.8$): (a1) net heat current, (b1) inner temperature profile for site 4, and (c1) leakage currents. Bottom ($\gamma_{I}=4.0$): (a2) net heat current, (b2) inner temperature profile for site 4, and (c2) leakage currents. Parameters which are used : $N=6$, $T_{1}=0.5$, $T_{N}=0.25$, $\gamma_{L}=\gamma_{R}=0.2$.}\label{fig2}
\end{figure*}
\section{Conclusions}
It is a very much testing experience to simulate quantum heat transport and it is successfully displayed in the present paper using QSCRM. Most of the anticipated results of heat transport in molecular junctions can be captured using this phenomenological QSCRM. The dependence of the heat transfer on the chain length, scattering rate, and temperature are correctly emulated through this QSCRM. Since this method can generate easily the classical and the quantum linear-response limits, one can use this method as a testing bed to interpolate between the ballistic and diffusive regimes. Our results also handy to reproduce the out of equilibrium condition, quantum effects and anharmonicity through this simple phenomenological model.\\

\indent
Furthermore, the present study exhibits the effect of different environmental spectrum and the phonon mismatch on the heat transport in MJs. Experimentally observed ``turnover" phenomena is correctly reproduced and the effect of different environment on such turnover effect is also demonstrated. As the QSCRM is capable to reproduce the coherent to diffusive transition, one can utilize this method to study  the elastic and ballistic transport in short alkane chains as well as the diffusive transport in some specific oligomers (polyethylene glycol). The in-situ local temperature evolution is helpful in study local temperature profile and the thermalization processes in MJs.\\
\begin{figure*}[t!]
  \centering
  \includegraphics[width=6cm]{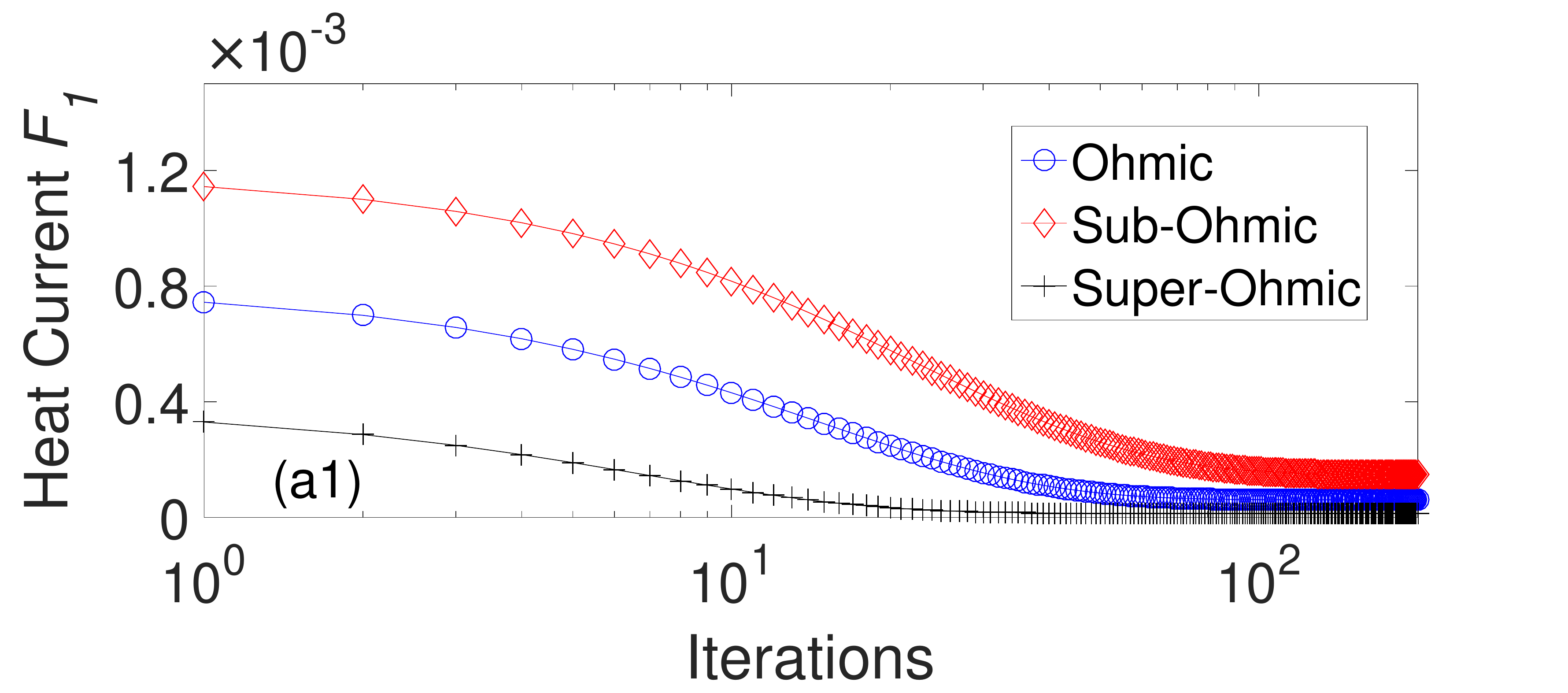}
  \includegraphics[width=6cm]{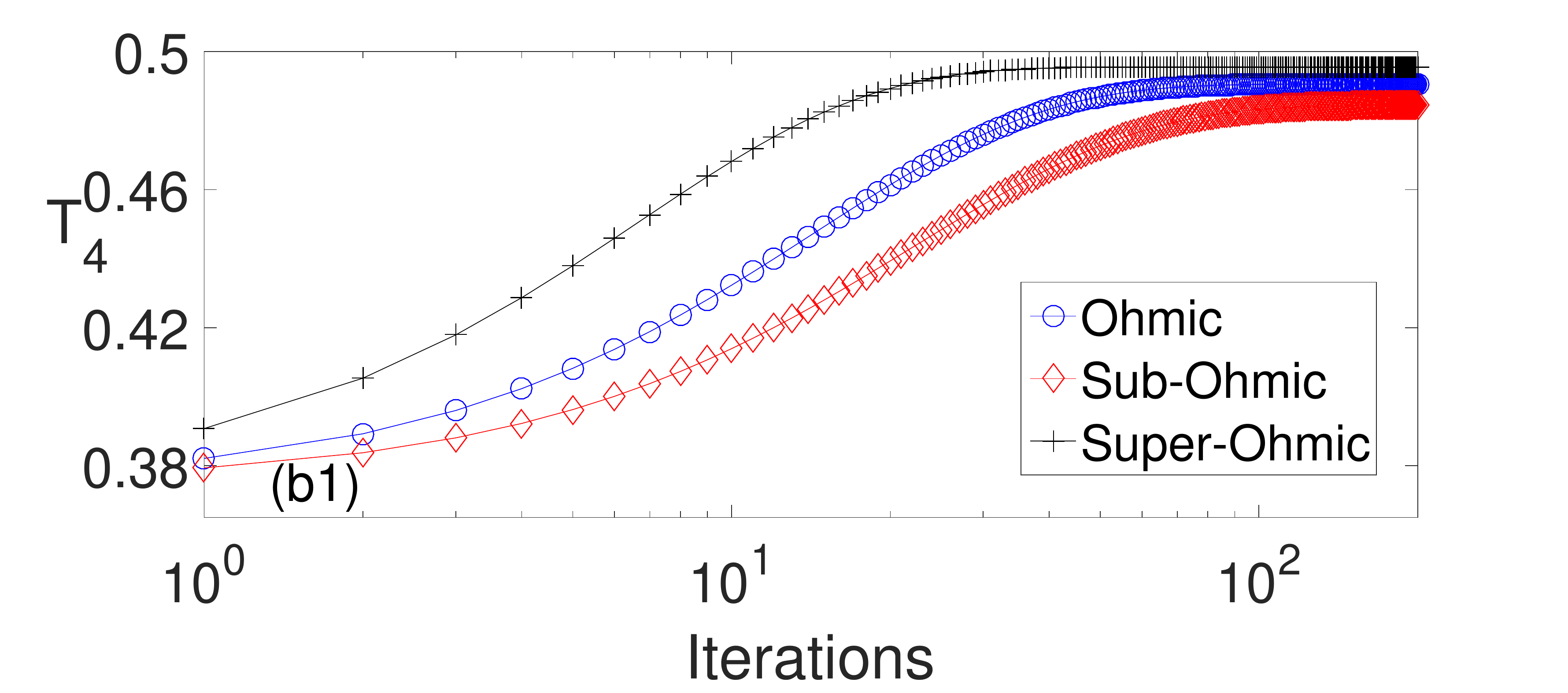}
  \includegraphics[width=6cm]{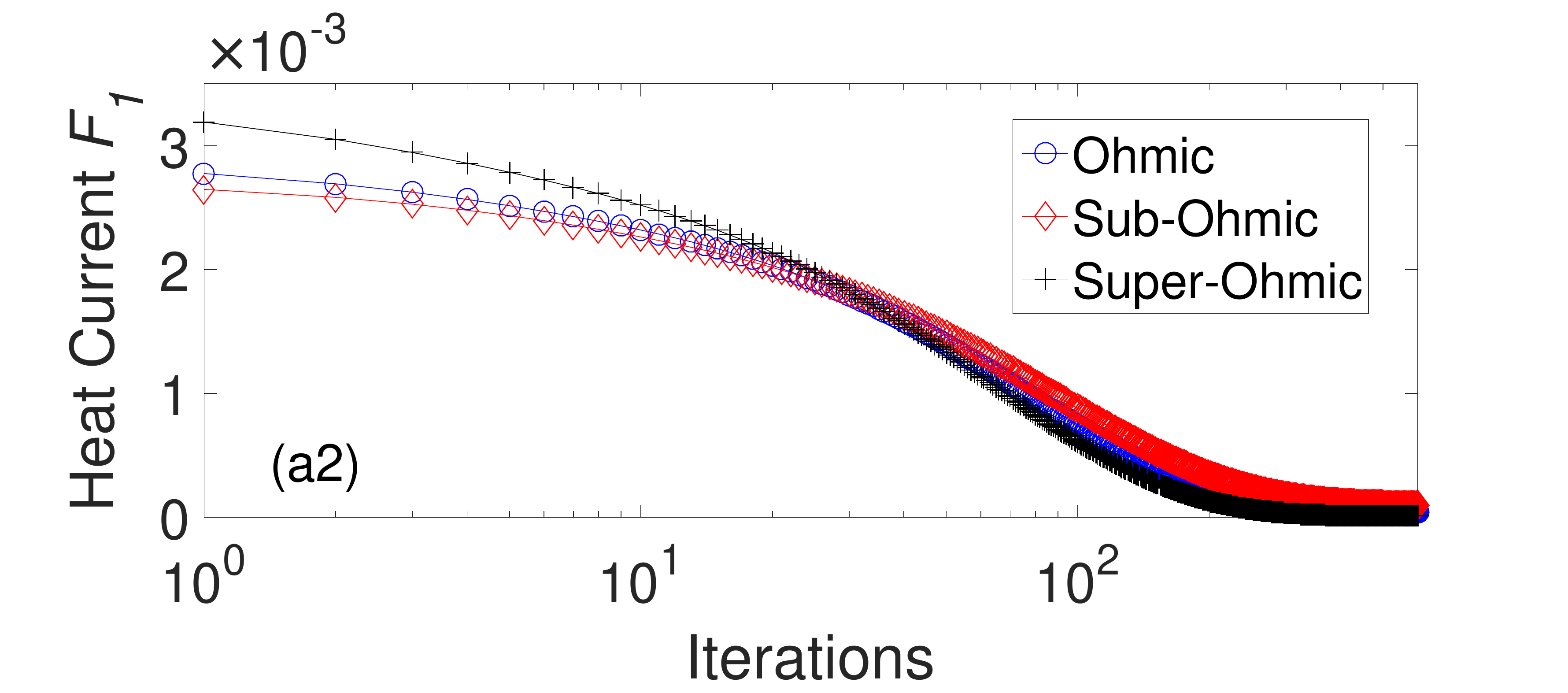}
  \includegraphics[width=6cm]{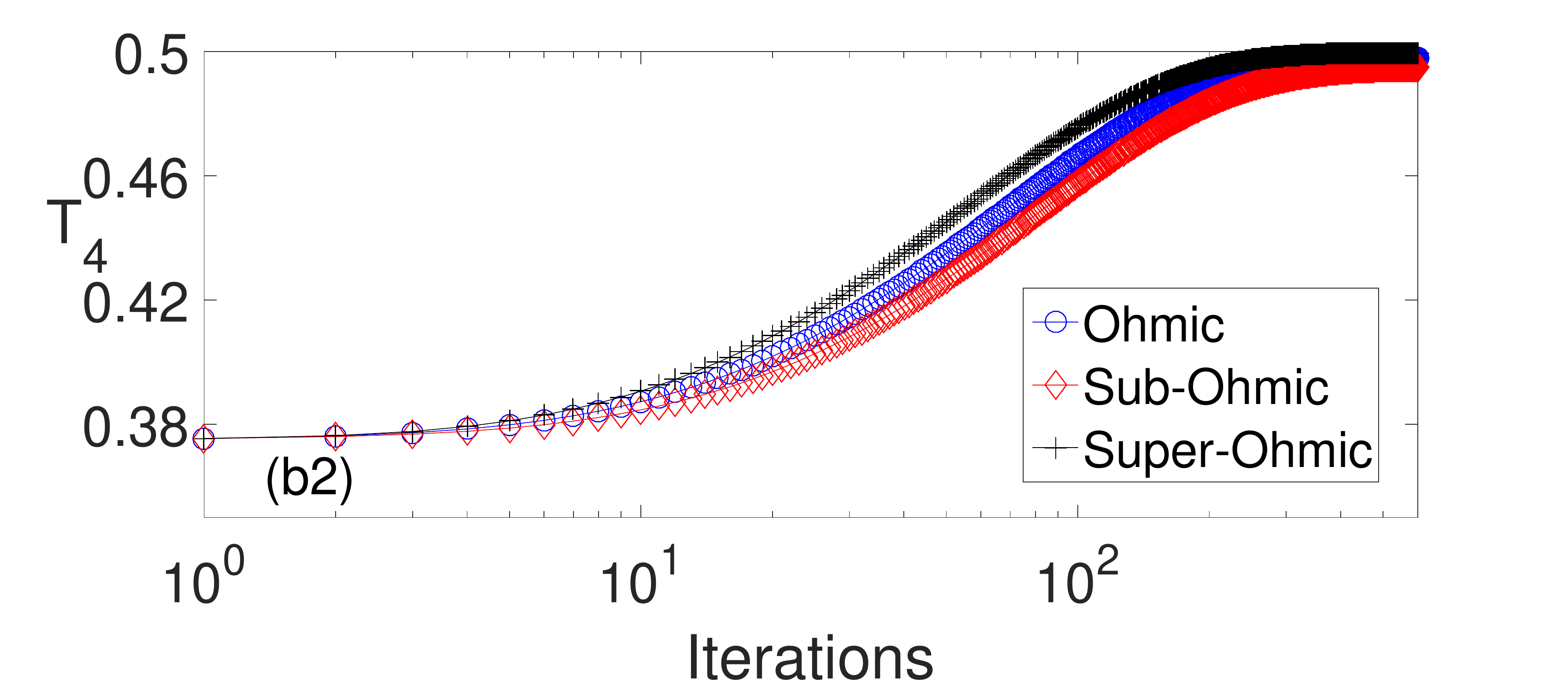}
  \caption{Convergence of the QSCR method with the increasing number of iterations using structured baths with exponential cut-off. Top ($\omega_{d,I}=0.1$): (a1) net heat current, and (b1) inner temperature profile for site 4.
  Bottom ($\omega_{d,I}=1$): (a2) net heat current, and (b2) inner temperature profile for site 4. Parameters which are used are : $N=6$, $T_{1}=0.5$, $T_{N}=0.25$, $\gamma_{L}=\gamma_{R}=0.2$, $\gamma_{I}=4.0$, $\omega_{d,L}=1$ and $\omega_{d,R}=0.1$.}\label{fig3}
\end{figure*}
\indent
The QSCR method can be implemented to study heat transport in large disordered molecules like DNA, proteins, and amorphous polymers. Using QSCRM one can go beyond one-dimension to study heat transport in realistic structures. In future one can use the QSCR method in combination
with atomistic Green’s function techniques to test the survival of quantum coherent effects in
single-molecule phononic conductance [30,31,32].\\
\appendix*
\section{ Numerical procedure}
Numerical procedure to solve the QSCRM equations are described in details in Refs. [51,53]. For completeness we review here some basics of the method discussed in Refs. [51,53]. The roots of Eq.\,(\ref{eq:10}) is obtained using the Newton-Raphson technique [78]. The root $r$ of a given well-behaved function $f(x)$ satisfying $f(r)=0$ can be obtained by iteratively solving
\begin{equation}\label{eq:13}
  x_{k+1}=x_{k}-\frac{f(x_{k})}{f^\prime(x_{k})},
\end{equation}
where $x_{0}$ is the initial guess value and $f^\prime(x_{k})$ is the first derivative of $f(x)$ at $x_{k}$. We choose the average temperature of the system, $T_{i}^{(0)}=(T_{1}+T_{N})/2, i=2,3,..., N-1$, as the initial guess for the profile. The temperature of the SC reservoirs are corrected at each iteration $k$ using
\begin{equation}\label{eq:14}
  T_{i}^{(k+1)}=T_{i}^k-\sum_{j=2}^{N-1}(D^{-1})_{i,j}F_{j}(T^{(k)}).
\end{equation}
Here the vector $T^{(k)}$ is the temperature profile after the $k$th iteration and $D$ is the Jacobian matrix with elements $D_{i,j}=\partial F_{i}/\partial T_{j}$.\\
\indent
To verify the convergence we examine the three quantities: (1) The temperature of the SC bath should approach a fixed value between $T_{1}$ and $T_{N}$. (2) The net heat current flowing through the system, $F_{1}$ should remain fixed. (3) Net exchange of heat current between the SC baths and chain should be vanishingly small as compared to net heat current across the system.\\
\indent
One can identify two major sources of error in our method: (1)
we numerically integrate over frequency in Eq. (9),
by discretizing energies between a lower and upper
cutoffs. We ensure that considering a fine frequency step $\Delta\omega\sim 10^{-3}$ and a large energy
cutoff $\omega_c >> \omega_0$, with $\omega_0$ as the chain characteristic frequency, our results are robust against
$\Delta\omega$ and $\omega_c$; (2) We solve Equation (9) on a discretized temperature
grid. It can be mentioned that for a coarse grid the inner reservoirs’
temperatures may significantly deviate from the exact SC
values, and leakage occurs [28]. We can achieve the convergence by choosing
 a mesh fine enough such that $|F_l |/|F_1| < 10^{-5}$ for
$l =2,3,....(N-1)$. Further, the net energy exchange between the SC baths and the
system may accumulate to large values for large chain size, we also confirm in our
simulations that the incoming and outgoing current, $|F_1|$ and
$|F_N|$ (equal in principle), differ only by less than $0.1\%$.
In principle, we observe that one should adopt a delicate grid
for reaching a good accuracy for chains with $N > 10$.
We therefore consider a two-step procedure to improve
efficiency. In the first step we use a relatively rough grid with
$\delta T=\frac{T_1-T_N}{200}$, and the iterative procedure to solve Eq. (9) is
followed for convergence. This is in the sense that the temperature
profile stays fixed throughout the iteration procedure. However, these temperatures
still deviate from the optimal (SC) temperatures, and
significant leakage may take place. We denote by p the number
of iterations in this part. In the second step, we develop an individual mesh around each particle site by considering 200 elements around temperature sector $\lbrack T_l^{(p)}-\delta T,T_l^{(p)}+\delta T\rbrack$. With this individualized grid, we further iterate Eq. (9) for q more times to achieve optimum temperature profile. \\
\indent
The number of iteration required to achieve convergence depends on the temperature difference, chain length, average temperature, phonon scattering rate $\gamma_{I}$, and the cutoff frequency $\omega_{d}$. There will be more phonon scattering for large $\gamma_{I}$ thus making convergence more challenging which is demonstrated in Fig.\,\ref{fig2} for a chain of $N=6$ particles. We reach convergence after 50 iterations when $\gamma_{I}\sim\gamma_{L,R}$. But for $\gamma_{I}>\gamma_{L,R}$ the convergence reached after 200 iterations. Convergence also depends on the cutoff frequency of the SC bath. If the cutoff is low then convergence can be easily reached since only modes below cutoff frequency will suffer scattering effect. This point is
illustrated in Fig.\,\ref{fig3} where a chain of $N=6$ particles connected to two heat baths with different cutoff frequencies. We can obtain our results for chain size $N\sim 10$ within 2-3 hours. But, it takes almost 4 days to obtain results for large chain size say $N\sim 30$. \\
\begin{acknowledgments}
MB acknowledges the financial support from DST through the core project CRG/2020/001768.
\end{acknowledgments}

\end{document}